\documentstyle[eqsecnum,preprint,aps]{revtex}
\begin{document}
\draft
\title{HYPERSURFACE-INVARIANT APPROACH TO \\
COSMOLOGICAL PERTURBATIONS }

\author{D. S. Salopek$^1$ and J. M. Stewart$^2$}
\vspace{2pc}
\address{
{}$^1$Department of Physics, University of Alberta,
Edmonton, Canada T6G 2J1 }
\vspace{2pc}
\address{
{}$^2$Department of Applied Mathematics and Theoretical Physics,
University of Cambridge \\
Silver Street, Cambridge, CB3 9EW, UK }

\date{June 28, 1994}

\maketitle

\begin{abstract}
Using Hamilton-Jacobi theory,
we develop a formalism for  solving semi-classical cosmological
perturbations which does not require an
explicit choice of time-hypersurface.
The Hamilton-Jacobi equation for gravity interacting with matter
(either a scalar  or dust field)
is solved by making an Ansatz which includes all terms
quadratic in the spatial curvature. Gravitational radiation and
scalar perturbations are treated on an
equal footing. Our technique encompasses
linear perturbation theory and it also describes
some mild nonlinear effects.
As a concrete example of the method, we
compute the galaxy-galaxy correlation function as well as
large-angle microwave background fluctuations
for power-law inflation, and we compare with recent observations.

\end{abstract}

\pacs{\hfill DAMTP R94/25, \quad ALBERTA THY/20-94}


\widetext

\section{Introduction}

The freedom in choosing a time-hypersurface in general
relativity is sometimes viewed as a curse because
it leads to numerical problems, difficulties in quantization, etc..
However, for semi-classical analyses based on the Hamilton-Jacobi (HJ)
equation, it is actually a blessing.

Even in classical general relativity,
selection of a useful time foliation is often a difficult task. In
solving  Einstein's equations using the ADM form
(see, e.g., ref. \cite{MTW}),
arbitrary  choices must be made
for the lapse $N$ and the shift $N^i$ functions; these fields reflect
our liberty in choosing the time hypersurface as well as
the spatial coordinates. Hamilton-Jacobi theory provides
an elegant method of bypassing these very difficult decisions.

It is remarkable that the Hamilton-Jacobi equation for
general relativity refers neither to the lapse nor to the shift functions.
As a result, a solution for the generating functional $\cal S$ is valid
for {\it all} choices of the temporal and spatial coordinates.
The HJ equation is the natural starting point for
a hypersurface and gauge-invariant analysis --- it yields
a covariant formulation. It is analogous to the Tomonaga-Schwinger
equation that was applied successfully to quantum electrodynamics
\cite{Tomo.Schw}. However, solutions to the HJ equation are difficult
to obtain because one
must solve for the entire ensemble of evolving universes that is described by
{\it superspace}.

In a series of papers \cite{SS92} --\cite{PSS94}, we have
developed a systematic method of solving the HJ equation by
using an expansion in spatial gradients. Using the first few terms,
one can derive the nonlinear Zel'dovich approximation
\cite{Zel} and its higher order
corrections which describe the formation of pancake structures
in a dust-dominated Universe \cite{CPSS94} --\cite{SSC94}.
Causality is maintained in the relativistic
theory, and the final expressions are actually simpler than those
obtained from the Newtonian theory (see, e.g.,
Moutarde et al \cite{Moutarde91} and Buchert and Ehlers \cite{BE93}).

Moreover, Parry, Salopek and Stewart \cite{PSS94} derived a recursion relation
which enables one to compute the higher order terms
for the generating functional from the
previous orders. Two useful techniques were employed:
a conformal transformation of the 3-metric as well
as a line-integral in superspace. Deforming the contour of
integration corresponds to choosing an alternative time-integration
parameter. If the endpoints of integration were fixed, all
such contours gave identical results
provided the theory was invariant under reparametrizations
of the spatial coordinates. Time-reparametrization
invariance was closely related to spatial gauge-invariance.

However, in the semi-classical problems of interest to inflationary
cosmology, a finite number
of terms is insufficient. In this paper, we will consider
all terms which are quadratic in the curvature.
We will effectively sum an {\it infinite } number of terms
in the spatial gradient expansion.
Our procedure is analogous to that developed by Barvinsky and Vilkovisky
\cite{Barvinsky} in a very different context: the one-loop effective action
for gravity interacting with matter.
In this way, we recover the results of linear perturbation theory
considered during the 1982 Nuffield workshop
\cite{Nuff82} --\cite{Star82}, as well as those of
Mukhanov, Feldman and Brandenberger \cite{Mukhanov92}. Our formalism also
describes some mildly nonlinear effects
which  we hope to apply to stochastic inflation
\cite{Starobinsky} --\cite{Linde2}.
Although there have been some interesting proposals, the choice
of time-hypersurface in stochastic inflation still requires further
clarification \cite{SB2}. For example,
Linde {\it et al} \cite{Linde2} have pointed out that eternal inflation
may appear differently on various time-hypersurface slices.
A covariant formulation would be advantageous.

In Sec. II, we set forth the HJ equation and the momentum
constraint equation. We consider two case of physical interest:
(1) a scalar field in an inflationary
universe and (2) dust in a matter-dominated epoch,
with or without a cosmological constant.
One can factor out the effects of the long-wavelength background by
using a conformal transformation of the 3-metric.
We review the long-wavelength formalism using the concept of
{\it field-space diagrams} which provide a simple illustration
of the concept of hypersurface transformation.
These diagrams are similar in spirit to Minkowski diagrams
that proved useful in understanding special relativity.

In Sec. III, we suggest an Ansatz which is second order in the
spatial curvature.
Arbitrary coefficients appear that are functions of the matter field
and the spatial Laplacian operator. Substitution into the HJ equation
lead to two linear differential equations, which describe the
evolution of the scalar modes as well as the tensor modes of the
3-metric. For an inflationary cosmology, we choose initial conditions
that are consistent with the Bunch-Davies vacuum \cite{BUD78}.

There is growing interest in the gravitational waves
produced during inflation
because it was pointed out \cite{S92} that they
could provide a large part of the signal detected by
the DMR (Differential Microwave Radiometer) experiment
on the Cosmic Background Explorer (COBE) satellite \cite{WRIGHT92},
and yet be consistent with structure formation. In our approach,
density perturbations and gravitational waves are treated on an
equal footing. In Sec. V, we compute large angle microwave background
fluctuations for various inflationary models and we compare with
COBE's two-year data set \cite{BENNETT94}, \cite{GORSKI94}.

After the ground work had been laid by Dirac
(see, e.g., ref.\cite{DIRAC}),
the Hamilton-Jacobi equation for general relativity was first
written down by Peres
\cite{Peres} in 1962. For flat space-time, Kucha\v r
\cite{K70} solved for the semi-classical wave-functional describing
the ground state.
By making explicit gauge choices, Halliwell and
Hawking \cite{HH85} gave approximate results for the
wavefunctional during the inflationary epoch.
Salopek, Bond and Bardeen \cite{SBB89} quantized the system using
a Heisenberg formulation, and they computed the power spectra
numerically for numerous models utilizing either one or two scalar
fields. By expanding the action to second order in perturbations,
Mukhanov {\it et al} gave an alternative prescription for quantizing this
system in arbitrary gauges. Their final results were
elegant, but because they perturbed the lapse and
shift functions, their method was somewhat tedious.
Our method removes this unattractive feature and it
generalizes the method of Kucha\v r and  Halliwell \& Hawking.

(Units are chosen so that $c=8\pi G=8\pi / m_P^2= \hbar= 1$.
The  sign conventions of Misner, Thorne and
Wheeler \cite{MTW} will be adopted throughout.)

\section{The Hamilton-Jacobi equation for general relativity}

For a single scalar field $\phi$ interacting with gravity,
the HJ equation and the momentum constraint equation are:
\begin{mathletters}
\begin{eqnarray}
0={\cal H}(x)=&&\gamma^{-1/2} {\delta{\cal S}\over \delta\gamma_{ij}(x)}
{\delta{\cal S}\over \delta\gamma_{kl}(x)}
\left[2\gamma_{il}(x) \gamma_{jk}(x) - \gamma_{ij}(x)\gamma_{kl}(x)\right]
+ \nonumber\\
&& {1\over 2} \gamma^{-1/2}\left({\delta{\cal S}\over \delta\phi(x)}\right)^2
+\gamma^{1/2}V(\phi(x)) + \;
\left [ -{1\over 2}\gamma^{1/2}R
+{1\over 2} \gamma^{1/2}\gamma^{ij}\phi_{,i}\phi_{,j} \right ] \,  ,
\label{HJES} \\
0={\cal H}_{i}(x)=&&
-2\left(\gamma_{ik}{\delta{\cal S}\over \delta\gamma_{kj}(x)}
\right)_{,j} +
{\delta{\cal S}\over\delta\gamma_{kl}(x)}\gamma_{kl,i} +
 {\delta{\cal S}\over\delta\phi (x)} \phi_{,i}  \, .
\label{MCS}
\end{eqnarray}
\end{mathletters}
In the ADM formalism, the line element is written as
\begin{equation}
ds^2\equiv g_{\mu \nu} dx^\mu dx^\nu =
\left(-N^2+\gamma^{ij}N_iN_j\right)dt^2 + 2N_i dt\,dx^i +
\gamma_{ij}dx^i\,dx^j\ ,
\label{ADMdecomp}
\end{equation}
where $N$ and $N_i$ are the lapse and shift functions respectively,
and $\gamma_{ij}$ is the 3-metric.
In eq.(\ref{HJES}),  $R$ denotes the Ricci scalar of the 3-metric.
The object of chief importance is the generating functional
${\cal S}\equiv {\cal S}[\gamma_{ij}(x), \phi(x)]$.
For each universe with field configuration
$[\gamma_{ij}(x), \phi(x)]$, it assigns a number
which can be complex. The generating functional is
the `phase' of the wavefunctional in the semi-classical approximation:
\begin{equation}
\Psi \sim e^{i{\cal S}} \, . \label{WF}
\end{equation}
For the applications that we are considering, the prefactor
before the exponential is not very important, although
it has interesting consequences for quantum cosmology \cite{B93}.
The probability functional,
\begin{equation}
{\cal P} \equiv |\Psi|^2, \label{PROB}
\end{equation}
is just the square of the wavefunctional (see, e.g., ref. \cite{S92b}).
It is the focus of attention in cosmology
during inflation, and even during the matter-dominated era.
The Hamilton-Jacobi equation (\ref{HJES}) and the momentum
constraint (\ref{MCS}) follow, respectively, from the
$G^0_0$ and $G^0_i$ Einstein equations with the canonical
momenta replaced by functional derivatives of ${\cal S}$:
\begin{equation}
\pi^{ij}(x) = { \delta {\cal S} \over \delta \gamma_{ij}(x) } \, ,
\quad {\rm and} \quad
\pi^\phi(x) = { \delta {\cal S} \over \delta \phi(x) } \, .
\end{equation}
Eq.(\ref{HJES}) is the relativistic generalization of the
Newton-Poisson relation, whereas eq.(\ref{MCS}) demands
that the generating functional be invariant under
an arbitrary change of spatial coordinates
(see, e.g., Misner {\it et al} \cite{MTW}, p.1185).
If the generating functional is real,  the evolution of the 3-metric
for one particular universe is given by
\begin{mathletters}
\begin{equation}
\left(\dot\gamma_{ij}-N_{i|j}-N_{j|i}\right)/N =2\gamma^{-1/2}
\left(2\gamma_{jk}\gamma_{il}-\gamma_{ij}\gamma_{kl}\right)
{ \delta {\cal S} \over \delta \gamma_{kl} } \, ,
\label{evol.metric}
\end{equation}
whereas the evolution equation for the scalar field is
\begin{equation}
\left(\dot\phi-N^i\phi_{,i}\right)/N=\gamma^{-1/2}
{\delta {\cal S} \over \delta \phi} \, .
\label{evol.scalar}
\end{equation}
\end{mathletters}
Here $|$ denotes a covariant derivative with respect to the 3-metric
$\gamma_{ij}$. The lapse and shift function appear neither
in the HJ equation (\ref{HJES}) nor in the momentum constraint (\ref{MCS}).
Hence in HJ theory, all gauge-dependent quantities appear only in the
evolution equations, (\ref{evol.metric})  and (\ref{evol.scalar}), for the
metric and scalar field.

HJ methods can also be applied fruitfully
to systems of perfect fluids \cite{SS92}, \cite{SS93}. For example
the Hamiltonian  and momentum constraints for collisionless,
pressureless dust \cite{SCHUTZ} are given by:
\begin{mathletters}
\label{Sconstraints}
\begin{eqnarray}
0= {\cal H}(x)=&&\gamma^{-1/2} {\delta{\cal S}\over \delta\gamma_{ij}(x)}
{\delta{\cal S}\over \delta\gamma_{kl}(x)}
\left[2\gamma_{il}(x) \gamma_{jk}(x)
- \gamma_{ij}(x)\gamma_{kl}(x)\right] \nonumber \\
&& + \sqrt{1 + \gamma^{ij}\chi_{,i}\chi_{,j}}\,
{\delta{\cal S}\over\delta\chi (x)}
-{1\over 2}\gamma^{1/2}R + V_0 \, , \label{HJED} \\
0= {\cal H}_{i}(x)=&&-2\left(\gamma_{ik}{\delta{\cal S}\over
\delta\gamma_{kj}(x)}
\right)_{,j} +
{\delta{\cal S}\over\delta\gamma_{kl}(x)}\gamma_{kl,i} +
{\delta{\cal S}\over\delta\chi (x)} \chi_{,i} \, .
\label{MCD}
\end{eqnarray}
\end{mathletters}
A cosmological constant term denoted by $V_0$ has also
been included.
The dust field $\chi$ describes, for example,
cold-dark-matter particles. Its evolution equation is
\begin{equation}
\left(\dot\chi-N^i\chi_{,i}\right)/N=\sqrt{1+\chi_{|k}\chi^{|k}},
\label{evol.dust}
\end{equation}
whereas that for the metric is given by eq.(\ref{evol.metric}).
The 4-velocity $U^\mu$ of the dust is the 4-gradient of the
potential $\chi$,
\begin{equation}
U^{\mu} = - g^{\mu \nu} \chi_{,\nu} \; ,
\end{equation}
where $g^{\mu \nu}$ is the inverse of the 4-metric.

\subsection{Review of Long-Wavelength Solution}

\subsubsection{Scalar Field and Gravity}

For fields where the wavelength is long compared to
the Hubble radius, one may safely neglect second order
spatial gradients (terms within square brackets) in the
HJ equation (\ref{HJES}) for gravity interacting with
a scalar field. The resulting equation
has the trivial solution
\begin{equation}
{\cal S}^{(0)}[\gamma_{ij}(x), \phi(x) ]=
-2\int d^3x\,\gamma^{1/2}\, H\left[ \phi(x) \right] .
\label{zeroth}
\end{equation}
provided that the Hubble function $H \equiv H(\phi)$, a
function of a single variable, satisfies
the separated Hamilton-Jacobi equation of order zero \cite{SB1}:
\begin{equation}
H^2 = { 2 \over 3 } \left ( { \partial H \over \partial \phi } \right )^2
+ { V(\phi) \over 3 } \, .  \label{SHJES}
\end{equation}
Eq.(\ref{zeroth}) is also a solution of the momentum contraint
(\ref{MCS}) because the volume element $d^3x \gamma^{1/2}$
is invariant under reparametrizations of the spatial coordinates.
$H \equiv H(\phi)$ corresponds to the Hubble parameter in the
long-wavelength limit. We will examine in detail the special
case of inflation with an exponential potential \cite{LM85}
\begin{mathletters}
\begin{equation}
V(\phi)=V_0 \, \exp\left(-\sqrt{2\over p}\phi\right),
\label{s.f.potential}
\end{equation}
where $p$ is a constant that describes the steepness of the potential.
An exponential potential
arises naturally in the induced gravity model (see, e.g., ref. \cite{SBB89})
as well as in extended inflation \cite{KST90}.
For this case, one can find the exact general
solution \cite{SB1} of eq.(\ref{SHJES}). In particular,
the separated HJ equation of order zero has the attractor solution
\begin{equation}
H(\phi)= \left[{V_0\over 3 \left ( 1-1/(3p) \right)}\right]^{1/2}
\exp\left(-{\phi\over\sqrt{2p}}\right) \, .
\qquad {\rm (single\ scalar\ field)}
\label{zero.scalar}
\end{equation}
\end{mathletters}
It describes power-law inflation where the scale factors evolves
as $a(t) \propto t^p$, with $t$ being a synchronous time variable.
(In order that the scalar field convert its energy into radiation
and matter at the end of inflation, the scalar field should
have a minimum in its potential.
Hence, we interpret eq.(\ref{s.f.potential}) as describing the
asymptotic branch of the potential as $\phi \rightarrow - \infty$;
we will assume that microwave background fluctuations as well as
fluctuations for galaxy formation are generated on this branch.)
The separated Hamilton-Jacobi equation has been widely applied
in the reconstruction of the inflaton potential from cosmological
observations \cite{COPELAND93}.

\subsubsection{Dust Field and Gravity}

For the case of a dust field, the long-wavelength theory is found
by dropping $R$ and $\chi_{|i} \chi^{|i}$ in the HJ equation
(\ref{HJED}).
One can then attempt a solution analogous to eq.(\ref{zeroth})
except now $H \equiv H(\chi)$ is a function of $\chi$ satisfying
\begin{equation}
H^2 = -{ 2 \over 3 } { \partial H \over \partial \chi }
+ H_0^2  \, , \quad {\rm where} \quad H_0^2= {V_0 \over 3} \, .
\label{SHJED}
\end{equation}
The general solution is
\begin{mathletters}
\begin{equation}
H(\chi)= H_0 \, {\rm cotanh}\left[ {3 H_0 \over 2 }
(\chi -\widetilde \chi) \right ] \, ,
\end{equation}
where $\widetilde \chi$ is a homogeneous constant \cite{SS92}.
If the vacuum energy density $V_0$ is negligible, then
one recovers the Hubble parameter for a matter dominated universe
in the limit that the cosmological constant vanishes,
$H_0 \rightarrow 0$:
\begin{equation}
H(\chi)= { 2 \over 3 (\chi -\widetilde \chi) } \, .
\end{equation}
\end{mathletters}

The scale factor, $\gamma^{1/6}$,  defined to be the sixth root of the
determinant of the 3-metric,
\begin{equation}
\gamma^{1/6} = d(x)
\left ( {\partial H \over \partial \widetilde \chi} \right )^{-1/3} \, ,
\end{equation}
can be found by taking the derivative with respect to the parameter
$\widetilde \chi$ \cite{SS92}.  Here $d(x)$ is an arbitrary
function of the spatial coordinates.
For dust with cosmological constant, we see that
\begin{mathletters}
\begin{equation}
\gamma^{1/6} \propto d(x) \,
\left  [ \sinh \left ( {3H_0\over 2}
(\chi - \widetilde \chi) \right ) \right ]^{2/3},
\label{V0D}
\end{equation}
whereas for pure dust, we recover the result
\begin{equation}
\gamma^{1/6} \propto d(x) \, ( \chi - \widetilde \chi)^{2/3} \, .
\end{equation}
\end{mathletters}
Eq.(\ref{V0D}) is plotted in Fig.(2).

\subsection{Field-Space Diagrams}

The consequences of the long-wavelength approximation can be
illuminated using {\it field-space diagrams} where one plots the metric
variable versus the matter field.

\subsubsection{Scalar Field and Gravity}

For simplicity, we will assume that the shift function
vanishes. The evolution equations (\ref{evol.metric}),
(\ref{evol.scalar}) in the long-wavelength limit then become
\begin{mathletters}
\begin{equation}
\dot \gamma_{ij}/ N = 2 H(\phi) \, \gamma_{ij} \, ,
\end{equation}
\begin{equation}
\dot \phi/ N = -2 {\partial H \over \partial \phi} \, .
\end{equation}
\end{mathletters}
These can be simplified by defining the field,
\begin{mathletters}
\begin{equation}
\alpha(t,x)= \left [ \ln \gamma(t,x) \right ]/ 6 \, ,
\label{alphaA}
\end{equation}
and then letting
\begin{equation}
\gamma_{ij}(t,x) = e^{2 \alpha(t,x)} \; k_{ij}(x) \, ,
\label{alphaB}
\end{equation}
\end{mathletters}
where $k_{ij}(x)$ is independent of time (this is the most
interesting case) with det$(k)$=1. Hence one obtains
\begin{mathletters}
\begin{equation}
\dot \alpha/ N =  H(\phi) \, ,
\label{evm}
\end{equation}
\begin{equation}
\dot \phi/ N = -2 {\partial H \over \partial \phi} \, .
\label{evs}
\end{equation}
\end{mathletters}

We can integrate these equations by utilizing our
freedom in choosing time. If we
choose $t= \phi$ to be the time variable,
the lapse is defined through eq.(\ref{evs})
\begin{equation}
1/ N = -2 {\partial H \over \partial \phi} \, .
\end{equation}
Eq.(\ref{evm}) then becomes
\begin{equation}
{d \alpha \over d \phi} =  - { 1 \over 2 } { H(\phi) \over
{\partial H \over \partial \phi} } \, ,
\end{equation}
which involves only $\alpha$ and the independent variable $\phi$.
It may be integrated immediately leading to
\begin{equation}
\alpha(\phi, x)= \alpha_0(x) - { 1 \over 2 }
\int^\phi_0  d\phi^\prime \, { H(\phi^\prime) \over
{\partial H \over \partial \phi\prime} }
\label{traj0}
\end{equation}
where $\alpha_0(x)$ is an arbitrary function of $x$.
For the example of inflation with an exponential
potential, eq.(\ref{zero.scalar}), we obtain the trivial solution
\begin{equation}
\alpha(\phi, x)= \alpha_0(x) + \sqrt{p \over 2} \phi \, .
\label{traj1}
\end{equation}
In general, the metric variable $\alpha$ is inhomogeneous
on a surface of uniform scalar field. For a single scalar
field, $\zeta/ 3$ is defined to be the metric fluctuation
on a uniform $\phi$ slice:
\begin{equation}
\zeta/3 \equiv \Delta \alpha(\phi)
\equiv  \alpha(\phi, x_2) - \alpha(\phi, x_1)=
\alpha_0(x_2) - \alpha_0(x_1) \, ;
\label{zeta}
\end{equation}
it is the difference of $\alpha$
between two spatial points $x_2$ and $x_1$ on a time hypersurface of
uniform $\phi$ \cite{BST83}, \cite{SBB89}, \cite{SB1}.
It is independent of time.
For a single scalar field in the long-wavelength limit,
this is true in general (see eq.(\ref{traj0})), and not just
for the example of an exponential potential.

We plot the solutions eq.(\ref{traj1})
in the field-space diagram Fig.(1). Each curve
represents the evolution of the fields $(\phi, \alpha)$
for a given spatial point.
One may invert eq.(\ref{traj1}) to obtain $\phi$ as a function of
$\alpha$
\begin{equation}
\phi(\alpha, x)= -\sqrt{2 \over  p}  \alpha_0(x) + \sqrt{2 \over  p}
\alpha \, .
\label{traj2}
\end{equation}
We interpret this inversion as choosing a time-hypersurface
where $\alpha$ is uniform. In this situation, the scalar
field is inhomogeneous. The fluctuation in the scalar field between
two spatial points $x_2$ and $x_1$ is given by
\begin{equation}
\Delta \phi(\alpha) \equiv \phi(\alpha, x_2) -  \phi(\alpha, x_1)=
-\sqrt{2 \over  p} \; \left [ \alpha_0(x_2)- \alpha_0(x_1) \right ] =
-\sqrt{2 \over  p} \; \Delta \alpha(\phi)  \, .
\label{ualpha}
\end{equation}
It is related to the metric fluctuation through the
negative of the slope of the $\alpha$ versus $\phi$ trajectories.
The transformation from a surface of
uniform $\phi$ to one of uniform $\alpha$ is simply visualized
in the field-space diagram.

\subsubsection{Dust Field and Gravity}

A similar long-wavelength analysis can be repeated for the
dust field. The evolution equations are
\begin{mathletters}
\begin{equation}
\dot \alpha/ N =  H(\chi) \, ,
\label{evol.alpha}
\end{equation}
\begin{equation}
\dot \chi/ N = 1 \, .
\label{evd}
\end{equation}
\end{mathletters}
If $\chi$ is taken to be the time hypersurface,
then $N=1$, and eq.(\ref{evol.alpha}) may
be integrated using the same method that was employed for
a scalar field. More simply, one can just apply eq.(\ref{V0D}),
\begin{equation}
\alpha(\chi, x) = \alpha_0(x)+ { 2 \over 3} \ln
\sinh \left [ { 3 H_0  \chi \over 2}  \right ]  \, .
\label{sol.dust}
\end{equation}
We have set $\widetilde \chi=0$, which may always be arranged by
shifting $\chi$. We plot $\alpha$ as a function of $\chi$ in
the field-space diagram, Fig.(2).
Qualitatively it is the same as Fig.(1),
except that here the trajectories are not straight lines.

HJ formalism is particularly useful for
problems involving several fields. For example, in the long-wavelength
limit, we have shown how to solve exactly the case of two
fluids, a dust field
and a field describing blackbody radiation \cite{SS93}.
This solution describes adiabatic as well as isothermal perturbations.
In this paper, we will consider only one matter field at a time.
(Using a Taylor series expansion in synchronous time,
Comer {\it et al} \cite{COMER94} have
investigated various perfect fluids.)

These very simple considerations illustrate the role of time in
general relativity. The HJ formalism appears to indicate
that the most useful choices for the time hypersurface will be
one of the matter fields or the metric variable $\alpha$,
or some combination of the two. A hypersurface transformation amounts to
slicing a field-space diagram in a particular direction.
Such simple behavior is also manifested in the higher
order solutions to the HJ equation for general relativity
that we will consider in the next section.

\section{QUADRATIC CURVATURE APPROXIMATION FOR
GRAVITY PLUS SCALAR FIELD}

\subsection{Factoring Out the Long-Wavelength Background}

Before we attempt to solve the HJ equation (\ref{HJES}),
we subtract out the long-wavelength background from the
generating functional:
\begin{equation}
{\cal S} = {\cal S}^{(0)} + {\cal F} \ , \quad
{\cal S}^{(0)} = - 2 \int d^3x \gamma^{1/2} H(\phi) \ ,
\label{subtract}
\end{equation}
where $H$ satisfies the separated Hamilton-Jacobi equation (\ref{SHJES}).
The functional for fluctuations, $\cal F$, now satisfies
\begin{eqnarray}
&& -2 { \partial H \over \partial \phi}
{\delta {\cal F} \over \delta \phi} +
2 H \gamma_{ij} { \delta {\cal F} \over \delta \gamma_{ij} }
+ \gamma^{-1/2} {\delta{\cal F}\over \delta \gamma_{ij}(x)}
{\delta{\cal F}\over \delta\gamma_{kl}(x)}
\left[2\gamma_{il}(x) \gamma_{jk}(x) -
\gamma_{ij}(x)\gamma_{kl}(x)\right] \nonumber \\
&& + {1\over 2} \gamma^{-1/2}
\left({\delta{\cal F}\over \delta\phi(x)}\right)^2
-{1\over 2}\gamma^{1/2}R
+{1\over 2} \gamma^{1/2}\gamma^{ij}\phi_{,i}\phi_{,j}=0\ .
\label{HJE1}
\end{eqnarray}
Although superficially similar,  this step differs
in principle from the usual analysis of perturbations
on an homogeneous background \cite{HH85}, \cite{SBB89}, \cite{Mukhanov92}.
Here a long-wavelength
background is allowed, which is closely related to what
is done in stochastic inflation. The first term
${\cal S}^{(0)}$ is explicitly
invariant under reparametrizations of the spatial
coordinates --- gauge-invariance is manifestly
maintained. Moreover, no explicit choice of the time
parameter has been made.

The first line of eq.(\ref{HJE1}) may simplified if one introduces a
change of variables, $( \gamma_{ij}, \phi ) \rightarrow
(f_{ij}, u)$:
\begin{mathletters}
\begin{equation}
u = \int \  { d \phi \over -2 { \partial H \over \partial \phi } } \ ,
\quad f_{ij} = \Omega^{-2}(u) \, \gamma_{ij} \ , \label{changeA}
\end{equation}
where the conformal factor $\Omega \equiv \Omega(u)$ is defined through
\begin{equation}
{ d \ln \Omega \over d u} \equiv -2 { \partial H \over \partial \phi}
{ \partial \ln \Omega \over \partial \phi} = H \ . \label{changeB}
\end{equation}
\end{mathletters}
Functional derivatives with respect to the fields transform according to
\begin{equation}
{ \delta \over \delta \gamma_{ij} } = \Omega^{-2}(u)  \;
{ \delta \over \delta f_{ij} } \Big |_u \ , \quad
{\delta \over \delta \phi} = - {1 \over 2}
{1 \over { \partial H \over \partial \phi } }
\left ( \;{ \delta \over \delta u}  |_{f_{ij}} - 2 H f_{ij}
{ \delta \over \delta f_{ij} } \Big |_u \; \right  )  \ .
\end{equation}
In order to simplify the notation, we will henceforth
suppress the symbols $|_u$ and $|_{f_{ij}}$ which
denote the variables that are held constant during
differentiation. Utilizing the conformal 3-metric $f_{ij}$ instead
the original 3-metric $\gamma_{ij}$ is analogous
to using comoving coordinates rather than physical
coordinates in cosmological systems.
At long-wavelengths, a surface of uniform $u$ corresponds to
comoving, synchronous gauge because $N=1$ in eq.(\ref{evs}).
Even if $\phi$ (considered as a function of $H$) oscillates,
$u$ is monotonic. However, when one considers short-wavelength terms
associated with the functional ${\cal F}$ for fluctuations,
a surface of uniform $u$ no longer represents a synchronous gauge.

The HJ equation reduces to
\begin{eqnarray}
{ \delta {\cal F} \over \delta u }+
\Omega^{-3}(u) f^{-1/2} {\delta{\cal F}\over \delta f_{ij}}
{\delta{\cal F}\over \delta f_{kl}}
\left[2 f_{il} f_{jk} - f_{ij} f_{kl}\right] +
{ \Omega^{-3}(u) f^{-1/2} \over 8
\left ( { \partial H \over \partial \phi} \right )^2(u)} \,
\left[  {\delta{\cal F}\over \delta u } - 2 H(u) f_{ij}
{\delta{\cal F}\over \delta f_{ij}} \right]^2
=  { \delta {\cal S}^{(2)} \over \delta u }
\label{HJEP}
\end{eqnarray}
where the functional ${\cal S}^{(2)}$ is given by
\begin{mathletters}
\begin{equation}
{\cal S}^{(2)}[f_{ij}(x), u(x)]=\int d^3x f^{1/2} \left[ j(u) \widetilde R +
k(u) u^{;i} u_{;i}\right] \ . \label{S2A}
\end{equation}
{}From now on, a semi-colon will denote a covariant derivative with respect to
the conformal 3-metric, e.g., $u^{;i} \equiv f^{ij} u_{;j}$.
In addition,  $\widetilde R$ is the Ricci scalar of
$f_{ij}$. The $u$-dependent coefficients $j$ and $k$,
\begin{equation}
j(u)=\int_0^u {\Omega(u^\prime)\over 2}\;d u'\ + F ,\qquad
k(u)=H(u) \, \Omega(u) \ , \label{S2B}
\label{jkscalar}
\end{equation}
\end{mathletters}
where $F$ is an arbitrary constant,
were first derived in refs. \cite{SS92}, \cite{PSS94} in order that
the spatial gradient terms appearing in eq.(\ref{HJE1}),
\begin{equation}
{1\over 2}\gamma^{1/2}R
-{1\over 2} \gamma^{1/2}\gamma^{ij}\phi_{,i}\phi_{,j}=
{ \delta {\cal S}^{(2)} \over \delta u } \, ,
\end{equation}
may be expressed as a functional derivative with respect to $u(x)$
holding $f_{ij}(x)$ fixed.
The momentum constraint maintains the same form as before
but it is now expressed in terms of the new variables
$(f_{ij}, u)$.
\begin{equation}
{\cal H}_{i}(x)=-2\left(f_{ik}{\delta{\cal F}\over \delta f_{kj}}
\right)_{,j} +
{\delta{\cal F}\over\delta f_{kl}} f_{kl,i}
+ {\delta{\cal F}\over\delta u } u_{,i}=0 \ .
\label{Snewmomentum}
\end{equation}

\subsection{Integral form of HJ equation}

It will prove useful to work with an integral form
of the HJ equation. Before proceeding, we pause to consider
a simple illustration from potential theory.

\subsubsection{Potential Theory}

The fundamental problem in potential theory is: given a force
field $g^i(u_k)$ which is a function of $n$ variables $u_k$,
what is the potential $\Phi \equiv \Phi(u_k)$ (if it exists)
whose gradient returns the force field:
\begin{equation}
{\partial \Phi \over \partial u_i} = g^i(u_k) \quad ?
\end{equation}
Not all force fields are derivable from
a potential. Provided that the force field satisfies the
integrability relation,
\begin{equation}
0= {\partial g^i \over \partial u_j} - {\partial g^j \over \partial u_i} =
\left [{\partial  \over \partial u_j}, {\partial  \over  \partial u_i }
\right ] \, \Phi \, ,
\end{equation}
(i.e., it is curl-free),
one may find a solution which is conveniently expressed using a
line-integral
\begin{equation}
\Phi(u_k) = \int_C \sum_j dv_j \ g^j(v_l) \ .
\end{equation}
If the two endpoints are fixed, all contours return the same
answer. In practice, we will employ the simplest contour that
one can imagine: a line connecting the origin to the
observation point $u_k$. Using $s$, $0 \le s \le 1$,
to parameterize the contour, we define the variable
$v_l\equiv v_l(s, u_l)$,
\begin{equation}
v_l = s \, u_l \quad {\rm with} \quad  dv_l = ds \; u_l
\end{equation}
so the line-integral may be rewritten as
\begin{equation}
\Phi(u_k) = \sum_{j=1}^n \int_0^1 { ds \over s } v_j \ g^j(v_k) \ .
\end{equation}

Similarly, in solving for the generating functional, we
will employ a line-integral in superspace. The integrability
condition for the HJ equation follows from the Poisson bracket of the
constraints (Moncrief and Teitelboim \cite{MT72})
\begin{equation}
\{ {\cal H}(x^k), {\cal H}(x^{k^\prime}) \} =
[ \gamma^{ij}(x^k) {\cal H}_j(x^k)+
\gamma^{ij}(x^{k^\prime}) {\cal H}_j(x^{k^\prime}) ] \;
{ \partial \over \partial x^i } \delta^3  (x^k - x^{k^\prime}) \, .
\label{poisson}
\end{equation}
In fact, each contour of the line-integral corresponds to
a particular time-hypersurface choice.
Provided that the generating functional is invariant
under reparametrizations of the spatial coordinates,
(e.g., ${\cal H}_j$ vanishes in eq.(\ref{poisson})),
different time-hypersurface choices will lead to the same
generating functional. Hypersurface invariance is closely
related to gauge-invariance.

\subsubsection{Line-Integral in Superspace}

Following Parry, Salopek and
Stewart \cite{PSS94}, an  integral form of the
HJ eq.(\ref{HJEP}) may be constructed using a line-integral in superspace:
\begin{eqnarray}
{\cal F}[f_{ij}(x), u(x)] &&+
\int d^3x \int_0^1 { ds \over s } \ v \Omega^{-3}(v) f^{-1/2}
{ \delta{\cal F} \over \delta f_{ij} }
{ \delta{\cal F} \over \delta f_{kl} }
\left[2 f_{il} f_{jk} - f_{ij} f_{kl} \right] + \nonumber \\
&& \int d^3 x \int_0^1 { ds \over s }  { v \Omega^{-3}(v) f^{-1/2} \over 8
\left ( { \partial H \over \partial \phi}(v) \right )^2 } \,
\left[  { \delta {\cal F} \over \delta v } - 2 H(v) f_{ij}
{ \delta{\cal F} \over \delta f_{ij} } \right]^2
=  {\cal S}^{(2)}[f_{ij}(x), u(x)]
\label{IHJE}
\end{eqnarray}
In analogy with the example from potential theory,
we replace the index $j$ with the spatial coordinate $x$
and the finite sum $\sum_j$ with the integral $\int d^3x$.
The integrating parameter is again denoted by $s$, and
$v(x)= s u(x)$ represents a straight line in the superspace of
the {\it scalar field};
however, at each  spatial point, $f_{ij}(x)$ is held fixed
in the line-integral. As a result,
one may safely add an arbitrary functional of $f_{ij}$
to the right-hand side. No approximations have been made thus far;
eq.(\ref{IHJE}) is exact.

\subsection{ANSATZ}

For a spatial gradient expansion of ${\cal S}$
we have already shown how to set up a recursion relation
\cite{PSS94} which gives higher order terms from the
previous orders. Explicit solutions were given which were
accurate to fourth order in spatial gradients.
However, we now wish to investigate higher order terms.
Finding an explicit and exact expression which is valid to all orders
appears to be extremely difficult. Instead
we will examine an infinite subset consisting of all terms quadratic in the
Ricci curvature, but containing any number of spatial gradients.
Halliwell and Hawking \cite{HH85} used a similar approximation, although
our expansion will manifestly maintain hypersurface and gauge-invariance.
We hence make an Ansatz of the form,
\begin{equation}
{\cal F} = {\cal S}^{(2)}  + {\cal Q} \, , \quad {\rm with}
\label{ansatz1}
\end{equation}
\begin{equation}
{ \cal Q} = \int d^3x f^{1/2} \left [
\widetilde R \ \
\widehat S(u, \widetilde D^2) \ \widetilde R +
 \widetilde R^{ij} \ \widehat T(u, \widetilde D^2) \ \widetilde R_{ij}
- { 3 \over 8 }   \widetilde R \ \widehat T(u, \widetilde D^2) \ \widetilde R
\right ] \, ,
\label{ansatz2}
\end{equation}
where $\widehat S(u, \widetilde D^2)$ and $\widehat T(u, \widetilde D^2)$
are differential operators which are also functions of $u$.
$\widetilde D^2$ is the Laplacian operator with respect to the
conformal 3-metric, e.g.,
\begin{equation}
\widetilde D^2 \widetilde R \equiv \widetilde D^{;i} \widetilde D_{;i} \,
\widetilde R
\equiv   \widetilde R^{;i}{}_{;i}
= f^{-1/2} \left( f^{1/2} f^{ij} \widetilde R_{,j} \right )_{,i} \ .
\end{equation}
We refer to ${\cal Q}$ as the quadratic functional.
We interpret the operator $\widehat S$ for scalar
perturbations to be a Taylor series of the form
\begin{equation}
\widehat S(u, \widetilde D^2) = \sum_n S_n(u)( \widetilde D^2)^n  \, ,
\end{equation}
and similarly for the operator $\widehat T$ describing tensor fluctuations.
Note that $\widehat T$ is
sandwiched between two Ricci tensors, $\widetilde R_{ij}$, as well as
two Ricci scalars $\widetilde R$.
The full Riemann tensor $\widetilde R_{ijkl}$ does not appear in the Ansatz
because for three spatial dimensions because it may be written in terms of
the Ricci tensor (see, e.g., \cite{SSP93}).
Although the first and third terms in eq.(\ref{ansatz2}) may be
combined into a single one, the present form simplifies
the final evolution equations for $\widehat S$ and $\widehat T$
(see eqs.(\ref{RICCATI1S},b)).

\subsubsection{Order of Perturbation}

We now define the rules which determine the order
in perturbation of various terms.
By first order, we refer to terms such as $\widetilde R$ or
$\widetilde D^2 u$, or  $ \widetilde D^2 \widetilde  R$ or
$\widetilde D^4 u$, which vanish if the fields
are homogeneous; they may contain any number of
spatial derivatives. Quadratic terms are a product of
two linear terms. Some examples are
$u_{;i} u^{;i}$, $\widetilde D^2 u \, \widetilde D^4 \widetilde R$.
It should be clear how to determine other cases.

Other quadratic terms such as
$ (\widetilde D^2 u)^2$ and $ \widetilde R \widetilde D^2 u$,
could be included in the Ansatz (\ref{ansatz2}) as well.
However for the case of a single scalar field
as well as that for a single dust field, it may
be shown that they actually vanish. This is one of advantages
of utilizing the conformal transformation (\ref{changeA},b).
(This simplification does not occur for multiple matter fields,
and analogous terms would have to be included.)

All that is necessary now is to compute some functional
derivatives, and substitute into the integrated HJ
equation (\ref{IHJE}). It is useful to note that for
a small variation of the conformal 3-metric $\delta f_{ij}$
the corresponding change in the Ricci tensor is
\begin{equation}
\delta \widetilde R_{ij} = {1 \over 2 } f^{kl} \left [
\delta f_{li;jk} + \delta f_{lj;ik}- \delta f_{ij;lk}
- \delta f_{lk;ij} \right ] \, .
\label{small}
\end{equation}
In the integral form of the HJ equation, integration by parts
is permitted which simplies the analysis considerably.
(However, in its differential form (\ref{HJES})
one cannot simply discard total spatial derivatives.)
In addition, all cubic terms are neglected. For example
from eqs.(\ref{S2A},b) we find that
\begin{equation}
{ \delta {\cal S}^{(2)} \over \delta u}
- 2 H f_{ij} { \delta {\cal S}^{(2)} \over \delta f_{ij} }=
f^{1/2} \left ( { \Omega \over 2 } - Hj \right ) \widetilde R \, ,
\label{Q2D}
\end{equation}
where a term proportional to $u_{;i} u^{;i}$ has been dropped, otherwise
if (\ref{Q2D}) where squared, this term's contribution to the
integral HJ equation (\ref{IHJE})
would be cubic and higher which we are not considering here.
Other terms that appear are
\begin{eqnarray}
\int && d^3x \int_0^1 { ds \over s } \ v \Omega^{-3}(v) \,
f^{-1/2}
{ \delta{\cal S}^{(2)} \over \delta f_{ij} }
{ \delta{\cal S}^{(2)} \over \delta f_{kl} }
\left[2 f_{il} f_{jk} - f_{ij} f_{kl} \right]
= \nonumber \\
&& \int d^3x \int_0^1 { ds \over s }  \,
f^{1/2} \; 2 v \Omega^{-3}(v) \, j^2(v)  \left
( \widetilde R^{ij} \widetilde R_{ij} - { 3 \over 8 } \widetilde R^2
\right )   \, .
\end{eqnarray}
The order of some operators may be changed if the
commutator is an undesired higher order term, e.g.,
\begin{equation}
\int f^{1/2} \widetilde R \widetilde D^2 \ \widehat T \ \widetilde R =
\int f^{1/2} \widetilde R \ \widehat T \ \widetilde D^2  \widetilde R \, .
\end{equation}
Hence, $ \widetilde D^2$ effectively commutes with any
function of $u$. For a similar reason, one may permute the
order of spatial differentiation with impunity,
$ \tilde D_{;i} \tilde D_{;j} \sim  \tilde D_{;j} \tilde D_{;i}$,
because the additional Riemann tensor term would be of higher order.
As a result, we find
\begin{eqnarray}
&&\int  d^3 x \int_0^1  { ds \over s } v { \Omega^{-3}(v) f^{-1/2} \over 8
\left ( { \partial H \over \partial \phi}(v) \right )^2 } \,
\left[  { \delta {\cal F} \over \delta v } - 2 H f_{ij}
{\delta {\cal F} \over \delta f_{ij} } \right]^2 =  \int d^3x f^{1/2}
\widetilde R \int_0^u dv \Bigg\{ { 1 \over 8 \Omega^3(v)
\left ( { \partial H \over \partial \phi}(v) \right )^2 }  \nonumber\\
&&
\left [
64 H^2(v) \ \widehat S^2 \widetilde D^4 +
16 H(v) \left( { \Omega(v) \over 2 } - H(v) \, j(v) \right )
\widehat S \widetilde D^2 +
\left( { \Omega (v) \over 2 }  - H(v)j(v)\right)^2\
\right ]
\Bigg \} \widetilde R   \, ,
\end{eqnarray}
and that
\begin{eqnarray}
\int d^3x \int_0^1 &&{ ds \over s } \ v \Omega^{-3} f^{-1/2}
{ \delta{\cal F} \over \delta f_{ij} }
{ \delta{\cal F} \over \delta f_{kl} }
\left[2 f_{il} f_{jk} - f_{ij} f_{kl} \right] = \nonumber \\
&& \int d^3 x f^{1/2}  \left \{
\widetilde R^{ij} \int_0^u dv \Omega^{-3}
\Bigg [ 2 j^2 + 4 j \ \widehat T \widetilde D^2 + 2 \ \widehat T^2
\widetilde D^4 \right ]
\widetilde R_{ij} \nonumber \\
&& - { 3 \over 8 } \widetilde R \int_0^u dv \Omega^{-3}
\left [ 2 j^2 + 4 j \ \widehat T \widetilde D^2 + 2 \ \widehat T^2
\widetilde D^4 \Bigg ]
\widetilde R  \right \}   \, .
\end{eqnarray}

Collecting terms, the integral form of the HJ equation
(\ref{IHJE}) gives:
\begin{eqnarray}
0= \int && d^3 x f^{1/2}  \Bigg \{
\widetilde R  \left [ \widehat S + \int_0^u dv
{ 1 \over 8 \Omega^3(v)
\left ( {\partial H \over \partial \phi} (v)\right )^2 } \
\left ( { \Omega(v) \over 2 } - H(v)j(v) +
8 H(v) \widehat S \widetilde D^2 \right )^2
\right ] \widetilde R  + \nonumber \\
\widetilde R^{ij} &&
\left [ \widehat T + \int_0^u dv { 2 \over \Omega^3(v)}
( j(v)+ \widehat T \widetilde D^2 )^2
\right ] \widetilde R_{ij}  -{ 3\over 8}
\widetilde R
\left [ \widehat T + \int_0^u dv { 2 \over \Omega^3(v)}
( j(v)+ \widehat T \widetilde D^2 )^2
\right ] \widetilde R \Bigg \}   \, .
\label{collect}
\end{eqnarray}
Each coefficient must vanish separately, leading to a
pair of uncoupled integral equations. Taking the derivative
with respect to $u$ for each, one obtains the
differential equations
\begin{mathletters}
\begin{eqnarray}
0&&= { \partial \widehat S \over \partial u} + { 1 \over 8 \Omega^3
\left ( { \partial H \over \partial \phi} \right )^2  } \
\left ( { \Omega \over 2 } - Hj + 8 H \widehat S \widetilde D^2 \right )^2
\, ,
\label{RICCATI1S} \\
0&&= { \partial \widehat T \over \partial u} + { 2 \over  \Omega^3 }
\left ( j + \widehat T \widetilde D^2 \right )^2
\label{RICCATI2S}   \, .
\end{eqnarray}
\end{mathletters}
Since one may add an arbitrary functional of $f_{ij}$ to
the integral HJ eq.(\ref{IHJE}), these first order, ordinary differential
equations are equivalent to the pair of integral equations
arising from eq.(\ref{collect}). These nonlinear differential
equations are of the Riccati type. They represent a tremendous
simplification over the original HJ equation (\ref{HJES}).

\subsection{SOLVING RICCATI EQUATIONS}

As is well known, the Riccati equations, (\ref{RICCATI1S}) and
(\ref{RICCATI2S}), may be reduced to linear ordinary differential equations.
To this aim, we define the Riccati transformation
\begin{equation}
w\equiv w(u, \widetilde D^2) \, , \quad
y\equiv y(u, \widetilde D^2) \, ,
\end{equation}
through
\begin{equation}
\widehat S = { \Omega^3 \left ( { \partial H \over \partial \phi } \right )^2
\over 8 H^2 \widetilde D^4} \
{ 1 \over w} { \partial w \over \partial u} +
{   2 Hj - \Omega \over 16 H \widetilde D^{2}  } \, ,
\label{ST}
\end{equation}
\begin{equation}
\widehat T = { \Omega^3 \over 2 \widetilde D^4} \
{ 1 \over y} { \partial y \over \partial u} - \widetilde D^{-2} j \, ,
\label{TT}
\end{equation}
which lead to the desired result:
\begin{eqnarray}
0 =&& { \partial^2 w \over \partial u^2} +
\left \{ 3 H(u) + 2{ \partial \over \partial u} \left [ \ln \left (
{ 1 \over H } { \partial H \over \partial \phi }  \right ) \right ]
\right \}
{ \partial w \over \partial u} - \Omega^{-2}(u) \widetilde D^2 w \, ,
\quad {\rm (scalar \; perturbations)} \label{scalar} \\
0 =&& { \partial ^2 y \over \partial u^2} + 3 H(u)
{\partial y \over \partial u} -  \Omega^{-2}(u) \widetilde D^2 y \, .
\quad {\rm (tensor\; perturbations)}
\label{tensor}
\end{eqnarray}
Equation (\ref{scalar}) describes the evolution of scalar perturbations
(density perturbations) of the metric, whereas eq.(\ref{tensor}) describes
tensor perturbations (gravitational waves).
We emphasize that no time choice has been made in deriving these
basic equations, although the scalar field parametrizes the evolution
because $u\equiv u(\phi)$ is the independent variable. These
are the fundamental equations of cosmological perturbations
that we will utilize in this paper.

If we set,
\begin{mathletters}
\begin{equation}
w = - { 1 \over 2 } { H \over {\partial H \over \partial \phi} } z \ ,
\label{htx}
\end{equation}
we obtain the equation for scalar perturbations that
was derived by Mukhanov {\it et al} \cite{Mukhanov92}
\begin{equation}
0={ \partial ^2 z \over \partial u^2}
+ 3 H(u)  {\partial z \over \partial u}+
(m^2_{eff} - \Omega^{-2} \widetilde D^2 ) z \, ,
\label{Mukh2}
\end{equation}
where the effective mass $m_{eff}$ is given by
\begin{equation}
m^2_{eff} = { \partial^2 V \over \partial \phi^2}
+ 2{ \partial H \over \partial u} \left [
3 - { 1 \over H^2} { \partial H \over \partial u}
+ 2 {\partial^2 \phi \over \partial u^2} \bigg/
\left (H  { \partial \phi \over \partial u} \right )
\right ] \, .
\label{Mukh3}
\end{equation}
\end{mathletters}
In deriving these equations, it is useful to note several
relationships between the scalar field $\phi$ and the
new variable $u$ which follow from eqs.(\ref{changeA}, b)
and (\ref{SHJES}):
\begin{mathletters}
\begin{eqnarray}
&&{d \phi \over du} = - 2 {\partial H \over \partial \phi} \ ,
\label{ida}\\
&&{d^2 \phi \over du^2} = -3 H {d \phi \over du} -
{ \partial V \over \partial \phi} \ , \label{idb} \\
&&{d^3 \phi \over du^3}\bigg/ {d \phi \over du}  = -3 {d H \over du}
- { \partial^2 V \over \partial \phi^2}
- 3H {d^2 \phi \over du^2} \bigg/ {d \phi \over du} \ . \label{idc}
\end{eqnarray}
\end{mathletters}
Eq.(\ref{idb}) is the well-known evolution equation for a
long-wavelength scalar field. It is identical to the evolution equation
in synchronous gauge ($N=1$) describing a homogeneous Friedman universe.

\subsubsection{Exact Solution}

Many exact solutions have been found for power-law
inflation (see, e.g., eq.(\ref{zero.scalar})).
Abbott and Wise \cite{AW84} have shown
that the evolution equation (\ref{tensor}) for tensor perturbations
can be solved exactly. In addition, Lyth and E. Stewart \cite{LS92} have
shown that the same solution may be applied to the
scalar perturbation eq.(\ref{Mukh2}). For example,
if we use $H(\phi)$ defined in eq.(\ref{zero.scalar}), we find that
\begin{equation}
{ 1 \over H} { \partial H \over \partial \phi} = -{ 1 \over \sqrt{ 2p} }
\end{equation}
is a constant, and the equation for scalar perturbations
eq.(\ref{scalar}) becomes
\begin{equation}
0 = { \partial^2 w \over \partial u^2} +
{3 p \over u} { \partial w \over \partial u} - u^{-2p} \widetilde D^2 w \, .
\quad {\rm (scalar \; perturbations)}
\label{nscalar}
\end{equation}
Here we have also used the fact that for power-law inflation, the
Hubble parameter and $\Omega$ are given by eqs.(\ref{changeA},b)
\begin{equation}
H(u) = { p \over u} \, , \quad {\rm and} \quad \Omega(u)= u^{p} \, .
\end{equation}
Eq.(\ref{nscalar}) is identical to that describing a {\it massless}
scalar field in a Friedman universe.
This is quite surprising since one would have naively
expected that the mass term $\partial^2 V / \partial \phi^2$
in eq.(\ref{Mukh3}) would
lead to damping of the fluctuations with wavelengths larger
than the Hubble radius but evidently
the additional terms cancel this effect.
In fact, the equation (\ref{nscalar}) for scalar perturbations
is  identical to the tensor evolution eq.(\ref{tensor}).
The solution for both of these equations can be expressed
in terms of a Hankel function of the first kind:
\begin{equation}
y(u, \widetilde D^2) = z(u, \widetilde D^2)=
\sqrt{2 \over p} w(u, \widetilde D^2)=
\sqrt{\pi u^{(1- 3p)} \over 4 |p-1| } \,   \, H_\nu^{(1)} \left [
{ \sqrt{ - \widetilde D^2} u^{1-p}  \over (1-p) } \right ] \,
, \quad {\rm with} \quad \nu= { (3p-1) \over 2( p-1) } \, .
\label{hankel2}
\end{equation}
The normalization of this solution is irrelevant since
a logarithmic derivative enters in the
Riccati transformation (\ref{ST}), (\ref{TT}).
However, in order to agree with the conventions of
Birrell and Davies \cite{BID78} we have assumed that
\begin{equation}
\Omega^3(u) \left ( y^*  { \partial y \over \partial u}-
y  { \partial y^* \over \partial u} \right )= i =
\Omega^3(u) \left ( z^*  { \partial z \over \partial u}-
z  { \partial z^* \over \partial u} \right )\, .
\label{norm}
\end{equation}
Note that the argument for the Hankel function
is negative, and it increases to zero as $u$ varies from
0 to $ \infty$. For an inflationary epoch where $p > 1$, it
describes a positive frequency solution as $u \rightarrow 0$,
e.g., when the wavelength of a  particular mode is far
within the Hubble radius,
\begin{equation}
y= z \sim { \left( -\tilde D^2 \right)^{-1/4} \over \sqrt{ 2 \Omega^2(u)} } \;
{\rm exp} \left( i \sqrt{ - \widetilde  D^2}
\int { du \over \Omega(u)} \right) \,  \quad (u\rightarrow 0) .
\end{equation}
As a result, eq.(\ref{hankel2}) corresponds to the ground state
wavefunctional at wavelengths shorter than the Hubble radius
(i.e., the Bunch-Davies vacuum \cite{BUD78}; see also ref. \cite{SBB89}).
However, when
the wavelength exceeds the Hubble radius, the state is no longer
in the ground state. (There is also an analogous solution
to eq.(\ref{nscalar})
which involves a Hankel function of the second kind
but it describes a negative frequency solution at short wavelengths --- it
does not describe a system which is initially in the ground state.)

\section{QUADRATIC CURVATURE APPROXIMATION FOR
GRAVITY PLUS DUST FIELD}

The quadratic curvature approximation used in Sec. III may
be applied to any field that is derived from an action principle.
We derive the scalar and tensor equations corresponding to a dust
field. Unfortunately one cannot define a ground state for the dust field
as was the case for
a scalar field. Hence the conditions at the beginning of the
dust-dominated era were generated at earlier times, such
as during the scalar field dominated epoch of inflation.

We follow the treatment given for a scalar field.
In analogy to eq.(\ref{subtract}), we express the generating functional
${\cal S} \equiv {\cal S}[\gamma_{ij}(x), \chi(x)]$ for dust and
gravity as the sum of
a long-wavelength  part ${\cal S}^{(0)}$ and a fluctuation part
${\cal F}$. Here, however, the Hubble function
$H\equiv H(\chi)$ is given by eq.(\ref{SHJED}).  The HJ equation
(\ref{HJED}) for dust then becomes
\begin{eqnarray}
&& {\delta {\cal F} \over \delta \chi}  +
2 H \gamma_{ij} { \delta {\cal F} \over \delta \gamma_{ij} }
+ \gamma^{-1/2} {\delta{\cal F}\over \delta \gamma_{ij}(x)}
{\delta{\cal F}\over \delta\gamma_{kl}(x)}
\left[2\gamma_{il}(x) \gamma_{jk}(x) -
\gamma_{ij}(x)\gamma_{kl}(x)\right] \nonumber \\
&& -{1\over 2} \gamma^{1/2} R  +
\left( \sqrt{1+ \chi^{|i} \chi_{|i} } - 1   \right )
\left( {\delta {\cal F} \over \delta \chi} -
2 \gamma^{1/2} { \partial H \over \partial \chi} \right )  \, .
\label{HJED1}
\end{eqnarray}
Since we cannot solve this equation exactly, we will retain
only those terms which are at most quadratic in perturbation:
\begin{eqnarray}
\left( \sqrt{1+ \chi^{|i} \chi_{|i} } - 1  \right )
\left( {\delta {\cal F} \over \delta \chi} -
2 \gamma^{1/2} { \partial H \over \partial \chi} \right )  \sim
{ 1 \over 2 } \chi^{|i} \chi_{|i}
\left( {\delta {\cal F} \over \delta \chi} -
2 \gamma^{1/2} { \partial H \over \partial \chi} \right )  \sim
- \gamma^{1/2}  \chi^{|i} \chi_{|i}
{ \partial H \over \partial \chi} \, .
\end{eqnarray}
We will also employ a conformal transformation of the 3-metric
$\gamma_{ij} \rightarrow  f_{ij}$
\begin{equation}
f_{ij} = \Omega^{-2}(\chi) \, \gamma_{ij} \, ,
\end{equation}
where $\Omega \equiv \Omega(\chi)$ is defined through
\begin{equation}
{d \ln \Omega \over d \chi} = H(\chi) \, .
\end{equation}
(For example, if the vacuum energy density $V_0$ vanishes
in eq.(\ref{SHJED}),
the Hubble function $H$ and the conformal factor can be written as
\begin{equation}
H= { 2 \over 3 \chi} \, , \quad  \Omega= \chi^{2/3} \, ,
\quad ( {\rm with} \; \tilde \chi=0) \, . )
\end{equation}
Hence, the HJ equation for dust becomes
\begin{equation}
\left.{ \delta {\cal F} \over \delta \chi }\right|_{f_{ij}}+
\Omega^{-3}(\chi) f^{-1/2} {\delta{\cal F}\over \delta f_{ij}}
{\delta{\cal F}\over \delta f_{kl}}
\left[2 f_{il} f_{jk} - f_{ij} f_{kl}\right] =
\left.{ \delta {\cal S}^{(2)} \over \delta \chi }\right|_{f_{ij}}
\label{HJEPD}
\end{equation}
where the functional ${\cal S}^{(2)}$ is given by
\begin{mathletters}
\begin{equation}
{\cal S}^{(2)}[f_{ij}(x), \chi(x)]=
\int d^3x f^{1/2} \left[ j(\chi) \widetilde R +
k(\chi) \chi^{;i} \chi_{;i}\right] \, ;
\end{equation}
once again, a covariant derivative with respect to
$f_{ij}$ is denoted using a semi-colon.
The $\chi$-dependent coefficients $j$ and $k$ are \cite{SS92},
\cite{PSS94},
\begin{equation}
j(\chi)=\int_0^\chi {\Omega(\chi^\prime)\over 2}\;d \chi'\ + F ,\qquad
k(\chi)=H(\chi) \, \Omega(\chi) \, ,
\label{jkdust}
\end{equation}
\end{mathletters}
where $F$ is a constant, so that
\begin{equation}
{1\over 2}\gamma^{1/2} R
+\gamma^{1/2} {\partial H \over \partial \chi} \chi^{|i} \chi_{|i}=
\left.{ \delta {\cal S}^{(2)} \over \delta \chi}  \right|_{f_{ij}}
\end{equation}
may be expressed as a functional derivative with respect to $\chi(x)$
holding $f_{ij}$ fixed.
Using a line-integral in superspace, we can construct the integral
form of this equation analogous to eq.(\ref{IHJE}):
\begin{equation}
{\cal F}[f_{ij}(x), \chi(x)] +
\int d^3x \int_0^1 { ds \over s } \ v \Omega^{-3}(v) f^{-1/2}
{ \delta{\cal F} \over \delta f_{ij} }
{ \delta{\cal F} \over \delta f_{kl} }
\left[2 f_{il} f_{jk} - f_{ij} f_{kl} \right]
=  {\cal S}^{(2)}[f_{ij}(x), \chi(x)] \, ,
\label{IHJED}
\end{equation}
where $v(x) = s \chi(x)$ represents a straight line in superspace.
This equation has a similar but simpler form than the case of a
scalar field. We make the analogous quadratic Ansatz as
in eq.(\ref{ansatz2}),
and we find that  the evolution equations for the scalar
and tensor operators,
$\widehat S \equiv \widehat S(\chi, \widetilde D^2)$ and
$\widehat T \equiv \widehat T(\chi, \widetilde D^2)$,  are
\begin{mathletters}
\begin{eqnarray}
0&&= { \partial \widehat S \over \partial \chi} \, , \label{scalard} \\
0&&= { \partial \widehat T \over \partial \chi} + { 2 \over  \Omega^3(\chi) }
\left ( j(\chi) + \widehat T \widetilde D^2 \right )^2 \, . \label{tensord}
\label{RICCATI2D}
\end{eqnarray}
\end{mathletters}
The solution to the scalar perturbation equation (\ref{scalard})
is trivial:
\begin{equation}
\widehat S \equiv \widehat S(\widetilde D^2) \quad
({\rm scalar \; perturbations})
\end{equation}
is an arbitrary functional of the conformal Laplacian
operator. It is independent of the dust field $\chi$.
Since dust never oscillates like a scalar field,
one cannot identify a ground state for the dust field.
The tensor perturbation equation (\ref{tensord})
may be solved using the same technique that was employed
for a scalar field. We first define $y$,
\begin{equation}
\widehat T = { \Omega^3 \over 2 \widetilde D^4} \
{ 1 \over y} { \partial y \over \partial \chi}
- \widetilde D^{-2} j(\chi) \, ,
\label{tdefd}
\end{equation}
which leads to a linear equation,
\begin{equation}
0= { \partial ^2 y \over \partial \chi^2} + 3 H(\chi)
{\partial y \over \partial \chi} -  \Omega^{-2}(\chi) \widetilde D^2 y \, ,
\quad {\rm (tensor\; perturbations)}
\label{tensord2}
\end{equation}
describing the evolution of the graviton in a universe with dust.

If the cosmological constant vanishes, we obtain the exact solution:
\begin{equation}
y= A \chi^{-1/2} J_{3\over2}
\left [ 3 \sqrt{ \widetilde D^2} \chi^{1/3} \right]  +
B \chi^{-1/2} J_{-{3\over2}}
\left [ 3 \sqrt{ \widetilde D^2} \chi^{1/3} \right] \, .
\label{exactd}
\end{equation}
The choice of the coefficients $A$ and $B$ will be
given in the next section. In fact, the forms for
$\widehat S$ and $\widehat T$ at the beginning of the
matter-dominated epoch are determined by a preceding
period of inflation with a scalar field.

\section{LARGE ANGLE MICROWAVE BACKGROUND FLUCTUATIONS AND
GALAXY CORRELATIONS}

We first show how to compute large angle microwave background
anisotropies arising from both the scalar and tensor
fluctuations of inflation. In principle this is not difficult
since it essentially amounts to expanding in spherical harmonics.
We first interpret the semi-classical
wavefunctional that was computed in the previous sections.

By maintaining gauge and hypersurface invariance
in the dynamical analysis, we have not made any
extraneous assumption which would typically complicate the results.
Consequently, the final equations for a scalar field,
(\ref{scalar}), (\ref{tensor}) and for a dust field,
(\ref{scalard}), (\ref{tensord2}) are of a very simple form.
However, many measurement processes actually choose
a specific time-hypersurface.
A simple example arises in special relativity:
when measuring the lifetime of a particle, one typically chooses the
rest frame of that particle. In an application of
high interest to cosmology, Sachs and Wolfe \cite{SW67}
derived the large angle microwave background anisotropy by utilizing
comoving, synchronous gauge (uniform $\chi$ slice).

\subsection{Interpretation of Semi-Classical Wavefunctional}

In earlier papers \cite{CPSS94} -- \cite{SSC94},
it was shown that the gradient expansion could
be used to compute nonlinear effects
in cosmology. In particular, we computed higher order
corrections to the Zel'dovich approximation.
However in computing microwave background anisotropies,
it is sufficient to consider a linear approximation;
we will now consider only a small deviation $h_{ij}$ of the conformal
3-metric $f_{ij}$ from flat space $\delta_{ij}$, where the probability
functional reaches its maximum:
\begin{equation}
f_{ij}(u, x) = \delta_{ij}+ h_{ij}(u,x) \, \quad {\rm with} \quad
h_{ij}(u, x)= \sum_{a=1}^6 \int { d^3 k \over (2 \pi)^3 }
\, e^{i \vec k \cdot \vec x} \, h_a(u, k) \, E_{ij}^{(a)}(k) \, .
\end{equation}
We have expressed  $h_{ij}$ as a sum of plane waves
with comoving wavenumber $\vec k$.
Furthermore, we also expanded it in a complete basis
of 6 symmetric,  polarization matrices $E_{ij}^{(a)}(k)$. For example, if the
comoving wavenumber is aligned
with the z-axis, we choose them to be:
\begin{mathletters}
\begin{eqnarray}
E^{(1)}_{ij}&= \, {\rm Diag} [ 1, -1, 0]  \, ,
\quad &{\rm (tensor)} \\
E^{(2)}_{12}&= E^{(2)}_{21} = 1 \, ,
\quad {\rm (all \; other \; components \; vanish)} \quad
&{\rm (tensor)} \\
E^{(3)}_{ij}&= \, {\rm Diag} [ 1, 1, 0]  \, ,
\quad &{\rm (scalar)} \\
E^{(4)}_{ij}&= {\rm Diag} [ 0, 0, \sqrt{2}] \, , \quad &{\rm (gauge)}\\
E^{(5)}_{13}&= E^{(5)}_{31}= 1 = E^{(6)}_{23} = E^{(6)}_{32} \, .
\quad {\rm (all \; other \; components \; vanish)} \quad
&{\rm (gauge)}
\end{eqnarray}
\end{mathletters}
The first two are traceless and divergenceless; they
correspond to tensor perturbations. The third describes scalar perturbations,
whereas the remainder are gauge modes. The matrices have been normalized
so that
\begin{equation}
E^{(a)}_{ij}   E^{(b)}_{ij} = 2 \delta^{ab} \, .
\end{equation}
In addition, one must respect the reality condition:
\begin{equation}
h_a(k)=h^*_a(-k) \, .
\label{reality}
\end{equation}
To  linear order, the Ricci tensor for the conformal metric
may be computed using eq.(\ref{small}) with $f_{ij} = \delta_{ij}$
and $\delta f_{ij} = h_{ij}$ to give
\begin{equation}
\tilde  R_{ij} = { 1 \over 2 } \left (
h_{li,j}{}^{,l} + h_{lj,i}{}^{,l} - h_{ij,l}{}^{,l}- h_{,ij} \right) \, .
\end{equation}
The probability functional arising from inflation is then given by the square
of the wavefunctional, eq.(\ref{PROB}):
\begin{mathletters}
\begin{equation}
|\Psi|^2[\gamma_{ij}(x), \phi(x)] =
{ \rm exp}- { 1 \over 2} \int { d^3 k \over (2 \pi)^3 }
\left[ |\beta_1(k)|^2 + |\beta_2(k)|^2+ |\beta_3(k)|^2 \right ] \, ,
\label{PROB1a}
\end{equation}
where
\begin{equation}
h_1(k) = { 1 \over \sqrt{ 2 k^4 T_I(u, -k^2)} } \, \beta_1(k) \, ,
\quad  h_2(k) = { 1 \over \sqrt{ 2 k^4 T_I(u, -k^2)} } \, \beta_2(k) \, ,
\quad  h_3(k) = { 1 \over \sqrt{ 16 k^4 S_I(u, -k^2)} } \, \beta_3(k) \, .
\label{PROB1b}
\end{equation}
\end{mathletters}
Here $\tilde D^2$ has been replaced by $-k^2$. In Fourier
space, $\widehat S$ and $\widehat T$ are complex numbers
which we expand into real and imaginary parts:
\begin{equation}
\widehat S(u,-k^2)= S_R(u,-k^2) + i S_I(u,-k^2) \, , \quad { \rm and} \quad
\widehat  T= T_R(u,-k^2) + i T_I(u,-k^2) \, ,
\end{equation}
with
\begin{equation}
S_I(u,-k^2) = {1 \over 16 k^4 }  \,
{ \left ( { \partial H \over \partial \phi } \right)^2  \over H^2 } \,
{ 1 \over |z|^2 } \, , \quad
T_I(u,-k^2) = { 1 \over 4 k^4 } \,
{1 \over |y|^2 } \, ,
\end{equation}
which follows from the Riccati transformation equations
(\ref{ST}), (\ref{TT}) and the normalization conditions (\ref{norm});
(we have assumed that $u$ is uniform (see below) and $j(u)$ is real).
Because the wavefunctional is invariant under reparameterizations
of the spatial coordinates, the gauge modes $h_{4}(k)$,
$h_{5}(k)$ and $h_{6}(k)$ are absent in eq.(\ref{PROB1a}).
Hence they are unrestricted, and
they may assume arbitrary values consistent with the reality condition
(\ref{reality}).
The three $\beta$'s are Gaussian random fields which satisfy the following:
\begin{equation}
< \beta_a(\vec k) \, \beta^*_b(\vec k) > =
(2 \pi)^3 \, \delta^3( \vec k - \vec k^\prime) \,
\delta_{ab} \, , \quad \beta_r(\vec k)= \beta^*_r(- \vec k) \, , \quad
a,b=1,2,3.
\end{equation}
The polarization matrices have been chosen so that the probability
functional (\ref{PROB1a}) is diagonal in $h_1(k)$, $h_2(k)$
and $h_3(k)$. (Using a canonical transformation in conjunction
with HJ theory, Langlois \cite{LANGLOIS93} has derived a reduced phase space
Hamiltonian; our approach differs from his in that we perform a phase space
reduction after finding a solution to the HJ equation.)

\subsubsection{Long-wavelength Fields from Power-law Inflation}

{\it Long-wavelength fields} are measurable and their evolution
was discussed in Sec. II.A. Recall that we defined $\zeta/3$ to be the
fluctuation in $ \alpha= (\ln \gamma)/ 6$ on
a comoving slice, eq.(\ref{zeta}).  By choosing $u\equiv u(\phi)$ to
be uniform in
eq.(\ref{PROB1a}), we see that for small deviations from flat
space that zeta is related to $h_3(k)$ through
\begin{equation}
\zeta(x) = { 3 \over 2 } \, \int { d^3 k \over (2 \pi)^3 }
\, e^{i \vec k \cdot \vec x} \, h_3(k)  \, .
\end{equation}
Since the gauge modes are arbitrary, we have chosen
$h_4= h_3/ \sqrt{2}$ in order that
\begin{equation}
h_3(u,k) \, E^{(3)}_{ij}(k) + h_4(u,k) \, E^{(4)}_{ij}(k) =
h_3(u,k) \delta_{ij}
\end{equation}
is proportional to the identity matrix.
It has become conventional to define the power spectrum through
\begin{equation}
{\cal P}_\zeta(k) \equiv { k^3 \over 2 \pi ^2} \,
\int d^3x  \; e^{-i \vec k \cdot \vec x}  < \zeta(x) \zeta(0) > \, ,
\label{convention}
\end{equation}
which leads to
\begin{equation}
{\cal P}_\zeta(k) \equiv { k^3 \over 2 \pi ^2} \, { 9 H^2 \over
4 \left ( {\partial H \over \partial \phi} \right)^2  } \, |z|^2 =
{ k^3 \over 2 \pi ^2} \, 9 |w|^2
\, .
\end{equation}
Loosely speaking, $[k^3 |w|^2/(2 \pi^2 )]^{1/2}$
can be interpreted as the
metric fluctuation, $\Delta \alpha(\phi) = \zeta/3$, on a
uniform $\phi$ slice (see eq.(\ref{zeta})), whereas
$[k^3 |z|^2/(2 \pi^2 )]^{1/2}$ is the fluctuation
in the scalar field, $\Delta \phi(\alpha)$, on a uniform $\alpha$
(uniform curvature) slice (see eq.(\ref{ualpha})). They are related
through the hypersurface transformation eq.(\ref{htx}); see
Sec.II.B and Fig.(1).
(In fact, Hwang \cite{Hwang93} has used uniform curvature
slices to provide an elegant derivation of the scalar
perturbation equation (\ref{Mukh2}).)
In DeSitter space where the scale factor varies exponentially
in synchronous time, $\Omega(u)= e^{Hu}$, where $H$ is a constant,
it is useful to note that that at long-wavelengths
\begin{equation}
\left[ { k^3\over (2 \pi^2 )}  |z|^2 \right ]^{1/2} =
\left[ { k^3\over (2 \pi^2 )}  |y|^2 \right ]^{1/2} = { H \over 2 \pi} \,
\end{equation}
is given exactly by the Hawking temperature \cite{BUD78}, \cite{SBB89}.
If inflation is not exactly DeSitter, this is approximately true
in which case it is useful to interpret $H\equiv H(k)$ as the Hubble
parameter at the time
that the comoving scale $k^{-1}$ crossed the Hubble radius
during inflation, $k/(H \Omega)\sim 1$, and
\begin{equation}
{\cal P}_\zeta(k) \sim  {9 \over 16 \pi^2 } \,
 H^4 / \left ( { \partial H \over \partial \phi} \right )^2 \, .
\end{equation}
For power-law inflation, we will compute this quantity exactly.

For the tensor modes, we define
$h_1(x)$ and $h_2(x)$ as well as their
corresponding power spectra, e.g.,
\begin{equation}
h_1(x) = \, \int { d^3 k \over (2 \pi)^3 }
\, e^{i \vec k \cdot \vec x} \, h_1(k)  \, ,
\end{equation}
with
\begin{equation}
{\cal P}_{h_1}(k)  \equiv { k^3 \over 2 \pi ^2} \,
\int d^3x  \; e^{-i \vec k \cdot \vec x}  < h_1(x) h_1(0) >
={ k^3 \over 2 \pi ^2} \, 2 |y|^2 \, .
\end{equation}
Using the exact solution eq.(\ref{hankel2}) in the long-wavelength limit,
$u \rightarrow \infty$,  we find
for power-law inflation that $y$ and $z$ are independent of time $u$:
\begin{equation}
|y| = |z|=  { 2^{\nu -1} \over \sqrt{\pi} } \,
{ \Gamma(\nu) \over \sqrt{p-1} }
\left( { k \over p -1} \right )^{ - (3p-1) \over 2(p-1) } \, .
\quad (u \rightarrow \infty)
\end{equation}
The long-wavelength power spectra for power-law inflation are
power-laws which are related to
each other through the steepness parameter $p$,
eq.({\ref{s.f.potential}):
\begin{mathletters}
\begin{equation}
{\cal P}_\zeta(k) = {\cal P}_\zeta(k_0) \,
\left ( { k \over  k_0 } \right )^{n_s-1} \, ,
\quad {\rm (long-wavelength)}
\label{poss}
\end{equation}
\begin{equation}
{\cal P}_{h_1}(k)={\cal P}_{h_2}(k)= { 4 \over 9 p} \,
{\cal P}_\zeta(k_0) \,
\left ( { k  \over  k_0 } \right )^{n_t-1} \, ,
\quad {\rm (long-wavelength)}
\label{post}
\end{equation}
\end{mathletters}
where, in the case of power-law inflation,
the spectral indices, $n_s, n_t$, for scalar and tensor
fluctuations  actually coincide
\begin{equation}
n_s = n_t = 1- { 2 \over ( p-1)} \, .
\end{equation}
${\cal P}_\zeta(k_0)$ is the value of the power spectrum in zeta
at some fiducial wavenumber scale $k_0$ which we will
choose later.  (It has been pointed out that inflation with
a cosine potential can also yield a power-law fluctuation
spectrum for scalar perturbations \cite{Adams93}.)

\subsubsection{Heating of the Universe}

At the end of inflation, the scalar field typically rolls to the minimum of
its potential where it oscillates and converts its energy into a thermal
bath consisting of radiation and matter. Because the coupling of
the scalar field
to matter is not well understood, several possibilities can arise
yielding either a low or high value of $T_{max}$, which we define to be
the maximum temperature reached immediately after inflation.
(Many models utilizing supersymmetry \cite{CDO93} give
values, $T_{max} \sim 10^{8}$GeV, whereas the Variable
Planck Mass model \cite{SBB89} is rather special in giving a very high result,
$T_{max} \sim 10^{15}$GeV.)
Fortunately, the amplitude of the metric at large wavelengths is
indifferent to the uncertainty
in $T_{max}$. (However, the present physical length of a
comoving scale is indeed sensitive to the value of the maximum temperature).
Numerical calculations for a simple phenomenological
model describing coupling of a single scalar field to radiation
and matter demonstrate that the amplitude of the metric fluctuation
$\zeta$ on a comoving slice remains constant \cite{SBB89}.
Essentially a result of momentum conservation,
the rest frame of the single scalar field
(uniform $\phi$) slice is coincident with the rest frame of radiation
which is identical to the rest frame of the matter (uniform $\chi$ slice)
at wavelengths larger than the Hubble radius.
Hence the fluctuations for structure formation arising from
inflation are typically {\it adiabatic}. (If there are
two scalar fields which are important during inflation, one
may also produce isocurvature perturbations which we will not consider here;
see, e.g., Sasaki and Yokoyama \cite{SY91}.)
Moreover, the function $y$ describing the tensor fluctuations
in eqs.(\ref{TT}) and (\ref{tdefd}) is continuous during the heating
process.

If we are only interested in large angle microwave background fluctuations
produced in the cold-dark-matter model, we may thus equate the probability
functionals on comoving time slices before and after heating of the Universe:
\begin{equation}
{\cal P}[ \gamma_{ij}| u= u_{heat} ]=
{\cal P}[ \gamma_{ij}| \chi= \chi_{heat}=0] \, .
\end{equation}
Hence for long-wavelength fields in the radiation and matter dominated
eras, the power spectra for zeta as well as that for tensor perturbations
are identical to those arising from inflation, eqs.(\ref{poss}) and
(\ref{post}). In other words, the scalar and tensor operators,
$\widetilde S$ and $\widetilde T$  are continuous on a comoving
time slice.

\subsection{Application of Sachs-Wolfe Formula}

It remains useful to continue employing comoving slices of uniform
$\chi$ since the phase transition where
radiation becomes uncoupled from matter occurs on a
slice where the temperature is uniform, $T= 4000K$, which
for adiabatic fluctuations, coincides with a uniform $\chi$
slice at large wavelengths (see, e.g., ref. \cite{SS93}). Hence by observing
the phase transition, one has effectively
{\it chosen} a very special time-hypersurface.
The Sachs-Wolfe \cite{SW67} formula yields the large angle temperature
anisotropy from a line integral over the perturbation in the
3-metric $h_{ij}$
computed in uniform $\chi$ gauge ({\it comoving, synchronous gauge}):
\begin{equation}
\Delta T(x) /T = -{ 1 \over 2 } e^k e^l \int_{t_1}^{t_2}
d \chi  \, { \partial h_{kl} \over \partial \chi }[ \chi, x(\chi)]
\label{SWF}
\end{equation}
The line integral traces the path $x(\chi)$ of a photon path from
the surface of last scattering to the present epoch; $e^i$ is
a unit vector giving the direction of the photon's velocity.
For angles of interest to COBE $(\alpha> 7^0)$, we are concerned with those
comoving scales that reenter the Hubble radius during the matter-dominated
era.

\subsubsection{Scalar Perturbations}

In comoving synchronous gauge, the scalar part of the metric evolves
according to
\begin{equation}
h^{(s)}_{ij}(\tau, x) = { 2 \over 3 } \zeta(x) \delta_{ij}
+ { 1 \over 15 } \tau^2 \, \zeta_{,ij}(x) \, ,
\quad {\rm (with}\; N=1, N^i=0).
\label{SM}
\end{equation}
(This may be derived most simply by using ref. \cite{SS92},
or by applying Sec.IV.)
Here we have defined conformal time $\tau \equiv \tau(\chi)$ to be
\begin{equation}
\tau = \int^\chi_0 { d \chi \over a(\chi)} = {2 \over H_0} \,
\left ( { \chi \over \chi_0} \right )^{1/3}
\, , \quad a(\chi) =
\left ( { \chi \over \chi_0} \right )^{2/3}, \quad
\chi_0= { 2 \over 3 H_0} \, ,
\end{equation}
where $H_0$ is the present value of the Hubble parameter which
we assume to be 50 km/s /Mpc which is consistent with measurements
of the Sunyaev-Zel'dovich effect by Birkinshaw {\it et al} \cite{B91}
(see also Lasenby \cite{L93}). In order to agree with our present
units of measurement, we have normalized the scale factor
$a(\chi) \propto \Omega(\chi)$ to unity at the present epoch $\chi_0$.
Integrating eq.(\ref{SWF}) twice by parts and retaining only those boundary
terms which are important for $\alpha > 2^0$, one finds that
the large-angle temperature anisotropy,
\begin{equation}
\Delta T(x)/ T= - \zeta(x)/15 \, ,
\end{equation}
in a flat, matter-dominated Universe is proportional to the value of
$\zeta(x)$ on the surface of last scattering which is a sphere
of radius $R=|x|= 11,700$Mpc. In this way, temperature fluctuations
are a probe of scalar fluctuations from inflation.

Since we are concerned with temperature anisotropies measured
on the celestial sphere, it is natural to employ a spherical harmonic
expansion. A plane wave  can be decomposed into orthogonal
spherical harmonic functions, $Y_{\ell m}$, and spherical Bessel
functions, $j_\ell$, through (see, e.g.,  ref. {\cite{jackson})
\begin{equation}
e^{i k \cdot x} = 4 \pi \sum_{\ell=0}^\infty \sum_{m=-\ell}^{m=\ell}
i^\ell j_\ell(kr) Y^*_{\ell m}( \Omega_x) \, Y_{\ell m}( \Omega_k) \, .
\label{add1}
\end{equation}
If $\delta$ denotes the angle between $\vec k$ and $\vec x$, then
the addition theorem relates the spherical harmonics with the
Legendre polynomial $P_\ell( {\rm cos}\delta )$:
\begin{equation}
P_\ell( {\rm cos} \delta) = { 4 \pi \over (2 \ell + 1) } \,
\sum_{m =-\ell}^{m=\ell} Y^*_{\ell m}( \Omega_x) \, Y_{\ell m}( \Omega_k) \, .
\label{add2}
\end{equation}
Hence, the plane wave expansion
\begin{equation}
\zeta(x) = \int { d^3 k \over (2 \pi)^3 }
\, e^{i \vec k \cdot \vec x}   \,
\left ( { 2 \pi^2 \over k^3} \,
{\cal P}_\zeta(k) \right )^{1/2} \, \beta_3(k) \, ,
\end{equation}
(which follows from the expression for the scalar power spectrum
eq.(\ref{convention}) of a Gaussian random field) implies that
the scalar contribution to the angular correlation
function $C_s(\alpha)$ can be expressed as a sum over the Legendre polynomials
$P_\ell( {\rm cos}\alpha )$ \cite{PEEBLES82}:
\begin{equation}
C_s(\alpha) \equiv < \Delta T(x) \Delta T(x^\prime) >_s =
\sum_{\ell=0}^\infty \,
<\Delta T_\ell^2>_s P_\ell( {\rm cos}\alpha ) \, ,
\end{equation}
\begin{mathletters}
\begin{eqnarray}
< \Delta T^2_\ell>_s\equiv && \left ( { T_\gamma \over 15 } \right )^2
(2 \ell +1) \int_0^\infty { dk \over k } \,
{\cal P}_\zeta(k) \, j^2_\ell(kR) \\
 && = A^2 \, { (2\ell +1) \over 5 } \,
{\Gamma( \ell + (n_s-1)/2) \over \Gamma(\ell + (5-n_s)/2)}
 \, { \Gamma((9-n_s)/2) \over \Gamma((3+n_s)/2) }, \\
A^2&& = T_\gamma^2 \, { \cal P}_\zeta(k_0)
 (k_0 R)^{1-n_s} \, { \pi  \over 45} \,  2^{n_s -4} \,
{ \Gamma(3-n_s) \over  \Gamma^2(2-n_s/2) } \,
{ \Gamma( (n_s+3) /2 )  \over {\Gamma( (9-n_s) /2)  } }
\end{eqnarray}
\end{mathletters}
where
$\alpha$ is the angle between the  two points, $x$ and $x^\prime$,
on the surface of last scattering; $T_\gamma = 2.736 \pm 0.017 K$ is
the mean background temperature \cite{GUSH90}.
${\cal P}_\zeta(k_0)$  was defined in eq.(\ref{poss}), and we
will assume that the fiducial wavenumber scale is $k_0=10^{-4}$Mpc.
(Integrals of various combinations of Bessel functions may be found
in Gradshteyn and Ryzhik \cite{GR80}).

\subsubsection{Tensor Perturbations}

The derivation of microwave anisotropies from tensor
perturbations is similar in principle to the scalar case. However
it technically more complicated because
the angular correlation function obtains contributions from
points within the surface of last scattering.

For primordial gravitational waves described by
eq.(\ref{post}), the metric for the tensor modes evolves according to
\begin{equation}
h_{ij}^{(t)}(\chi,x) = \int { d^3 k \over (2 \pi)^3 }
\, e^{i \vec k \cdot \vec x}  \,
\left( { 2 \pi^2 \over k^3}  {\cal P}_{h_1}(k) \right)^{1/2} \,
{ 3 j_1(k \tau) \over (k \tau) } \,
\sum_{a=1}^2 \beta_a(k) \, E^{(a)}_{ij}(k)  \, .
\end{equation}
during the matter-dominated era.  By applying a continuity argument
after the exit of inflation to the radiation and matter dominated
eras, we have determined the coefficients $A$ and
$B=0$ in eq.(\ref{exactd});
we have also chosen to rewrite eq.(\ref{exactd}) in terms of a
spherical Bessel function $j_1(k\tau)$ of order 1.
Using the Sachs-Wolfe formula eq.(\ref{SWF}), we find that
the temperature anisotropy is
\begin{equation}
\Delta T/ T = -{ 1\over 2} e^i e^j \int^{\tau_2}_{\tau_1} d \tau \,
\int { d^3 k \over (2 \pi)^3 }
\left( { 2 \pi^2 \over k^3}  {\cal P}_{h_1}(k) \right)^{1/2} \,
\, e^{i \vec k \cdot  \vec x(\tau)}
\, {\partial \over \partial \tau}
\left( { 3 j_1(k \tau) \over (k \tau) } \right)
\sum_{a=1}^2 \beta_a(k) \, E^{(a)}_{ij}(k)\, ,
\end{equation}
and the resulting two-point correlation function is
\begin{eqnarray}
< {\Delta T(x) \over T} {\Delta T(x^\prime) \over T} >_t && =
{ 1\over 4} e^i e^j  e^{\prime p} e^{\prime q}
\int^{\tau_2}_{\tau_1} d \tau \,
\int^{\tau_2}_{\tau_1} d \tau^\prime \,
\int { d^3 k \over (2 \pi)^3 }
{ 2 \pi^2 \over k^3}  {\cal P}_{h_1}(k) \,
e^{i \vec k \cdot  \vec x(\tau) -
i \vec k \cdot  \vec x^\prime(\tau^\prime) } \nonumber \\
&&{\partial \over \partial \tau}
\left( { 3 j_1(k \tau) \over (k \tau) } \right) \,
{\partial \over \partial \tau^\prime}
\left( { 3 j_1(k \tau^\prime) \over (k \tau^\prime) } \right) \,
\sum_{a=1}^2  \, E^{(a)}_{ij}(k) E^{(a)}_{pq}(k)\, ,
\label{tcorr}
\end{eqnarray}
where
\begin{equation}
x^i= \left( { 2 \over H_0} - \tau \right )\, e^i \, , \quad
x^{\prime i} = \left( { 2 \over H_0} - \tau^\prime \right )
e^{\prime i} \, ,
\end{equation}
describes the paths of the photons, with
$e^i = x^i / |x |$, $e^{\prime i} = x^{\prime i} / |x^\prime |$.
The values of conformal time at the present epoch and at the
time of decoupling, $(1+a)^{-1} \sim 1300$,
are, respectively,
$\tau_2 = 2/ H_0$ and $\tau_1= 2/(H_0 \sqrt{1300})$.
This expression may be simplified by noting that
\begin{equation}
\sum_{a=1}^2 E_{ij}^{(a)}(k) \, E_{pq}^{(a)}(k)
= \left [ \tilde \delta_{ip} \tilde \delta_{jq}+
\tilde \delta_{iq} \tilde \delta_{jp} -
\tilde \delta_{ij} \tilde \delta_{pq} \right ] \, , \quad
{\rm where } \quad
\tilde \delta_{ij} = \delta_{ij} - k_i k_j/ k^2 \, ,
\end{equation}
so that
\begin{equation}
e^i e^j e^{\prime p} e^{\prime q}
\sum_{a=1}^2 E_{ij}^{(a)}(k) \, E_{pq}^{(a)}(k)
= 2\left( {\rm cos} \alpha -
{\rm cos} \theta {\rm cos} \theta^\prime \right)^2
- \left( 1- {\rm cos}^2\theta \right ) \,
\left( 1- {\rm cos}^2\theta^\prime \right ) \, .
\end{equation}
Here $\theta$ denotes the angle between $\vec k$ and $\vec x$,
and analogously for $\theta^\prime$, and once again,
$\alpha$ is the angle between $x$ and $x^\prime$.
The various factors of cos$\theta$ (and cos$\theta^\prime$)
may be removed by several applications of the identity,
\begin{equation}
{\rm cos} \theta \;  P_\ell( {\rm cos} \theta)=
{ ( \ell + 1 ) \over ( 2 \ell + 1 ) } P_{\ell+1}( {\rm cos} \theta) +
{ \ell  \over ( 2 \ell + 1 ) } P_{\ell-1}( {\rm cos} \theta) \, ,
\end{equation}
in conjunction with the plane wave decomposition into
Legendre polynomials, eqs.(\ref{add1}) and (\ref{add2}).
The angular integrations in eq.(\ref{tcorr}) may be performed, and once
again the angular correlation function can written in
a series of Legendre polynomials,
\begin{equation}
C_t(\alpha)  = \sum_{\ell =0}^\infty \,
<\Delta T_\ell^2>_t P_\ell( {\rm cos}\alpha ) \, ,
\end{equation}
\begin{eqnarray}
<\Delta &&T_\ell^2>_t =   { 9 \over 4 } T_\gamma^2 \,
(\ell-1) \ell (\ell+1) (\ell+2) (2 \ell +1)  \\
&& \int_0^{2 k_{max} / H_0} dw w \;
{\cal P}_{h_1}\left( {H_0 w \over 2 } \right) \biggl \{
\int_0^{1 - (H_0 \tau_1)/2} \, dv \\
&&\left [ 3 { {\rm cos}\left ( w (1-v) \right ) \over w^3 (1-v)^3 }  +
 { {\rm sin}\left ( w (1-v) \right ) \over w^2 (1-v)^2 }
-3 { {\rm sin}\left ( w (1-v) \right ) \over w^4 (1-v)^4 } \right ] \\
&& \left [ { j_{\ell +2}( wv) \over ( 2 \ell +1) ( 2 \ell +3) } +
       2 { j_{\ell}( wv) \over ( 2\ell -1) ( 2 \ell +3) } +
       { j_{\ell  -2}( wv) \over ( 2 \ell +1) ( 2 \ell -1) } \right ]
\biggr \}^2
\end{eqnarray}
which was derived by Abbott and Wise \cite{AW84}
(see also Starobinsky \cite{STAR85}).
$k_{max}=10^{-2}{\rm Mpc}^{-1}$ corresponds to the Hubble
radius at matter-radiation equality. For angular scales that are measured
by COBE ($\alpha > 7^0$), the precise value of $k_{max}$
is not that important, since the spherical Bessel functions
$j_\ell(\omega v)$ provide a cutoff.

Since the scalar and tensor contributions are independent Gaussian
random fields, they add in quadrature
\begin{equation}
<\Delta T_l^2> = <\Delta T_l^2>_s + <\Delta T_l^2>_t  \, .
\label{sumcor}
\end{equation}
In Fig.(3), we have computed the relative contribution of the
tensor component
$<\Delta T_{\ell}^2>_t/ ( <\Delta T_{\ell}^2>_s + <\Delta T_{\ell}^2>_t)$
for various values of the spectral index
$n \equiv n_s = n_t =1 - 2/ (p-1)$. For smaller values of $n_s$,
it increases quite dramatically.

\subsection{Recent Observations}

The COBE DMR team has recently analyzed their 2-year data set
\cite{BENNETT94}. At the $68 \%$ confidence level,
they find that the spectral index \cite{GORSKI94} for scalar
perturbations is $n_s=1.10 \pm 0.32$, which is
consistent with the simplest
models of inflation models which yield $n_s < 1$. They
also determine that the root-mean-square temperature
anisotropy with dipole removed is $\sigma_{sky}(10^0)= 30.5 \pm 2.7 \mu K$
at the same level of confidence. The latter quantity,
$$
\sigma_{sky}^2(10^0) =
\sum_{l=2}^\infty <\Delta T_l^2> \,
{\rm exp}\left [  - l(l+1)/ 13.5^2 \right ] \, , \quad
\sigma_{sky}(10^0)= 30.5 \pm 2.7 \mu K \, ,
$$
is computed using the sum of the scalar and tensor
fluctuations eq.(\ref{sumcor}).
It determines the arbitrary normalization factor
appearing in long-wavelength power spectra, eqs.(\ref{poss},b).
The exponential factor corresponds
to a Gaussian window function with full width at half maximum of $10^0$.

However, COBE by itself cannot discriminate between tensor
and scalar fluctuations. One needs an additional experiment
to measure the scalar perturbations. Two proposals
have been suggested using either:
(1) galaxy clustering data  \cite{S92} or
(2) intermediate microwave background experiments
$1^0 <  \alpha < 2^0$ \cite{DAVIS92}.
We shall discuss only the first proposal here
since there are large variations in the  intermediate angle
observations \cite{INTERMED1} --\cite{INTERMED4}.

In Fig.(4), we have computed the power-spectra for zeta
that arises from the various power-law inflation models,
assuming that they account for COBE's measurement of $\sigma_{sky}(10^0)$.
In the limit that $n_s \rightarrow 1$ ($p \rightarrow \infty$)
gravitational waves do not contribute to COBE's signal,
and the power spectrum for zeta is the flat, Zel'dovich spectrum.
As $n_s$ decreases, gravitational waves are significant,
leaving a smaller contribution for the scalar perturbation
to $\sigma_{sky}(10^0)$. Hence at the fiducial wavenumber
$k_0=10^{-4}{\rm Mpc}^{-1}$
(scales probed by COBE), ${\cal P}_\zeta(k_0)$ decreases as $n_s$ decreases.
Moreover, the slope of the power spectrum becomes more negative as
$n_s$ decreases.

In comoving synchronous gauge (uniform $\chi$),
$\rho \gamma^{1/2}$ is independent of time and the linear density
perturbation at early times is
\begin{equation}
\delta(\tau, x) = \left ( \rho - \overline \rho \right)/ \overline \rho=
- { \tau^2 \over 30 } \, \zeta^{,l}_{,l}(x) \, ,  \quad
{\rm (early \; times)}
\end{equation}
which may be derived from the expression for the metric eq.(\ref{SM}).
In Fourier space this yields,
\begin{equation}
\delta(\tau, k) = { k^2 \tau^2 \over 30 } \, \zeta(k) \, .
\end{equation}
During the radiation-dominated era, density perturbations
oscillate and they damp because of the Hubble expansion.
This effect is described by the transfer function
$T(k)$, so that in the matter-dominated era the linear
density perturbation is given by
\begin{equation}
\delta(\tau, k) = { k^2 \tau^2 \over 30 } \, T(k) \, \zeta(k) \, .
\end{equation}
This requires an assumption for the dark matter, and we have
adopted the cold-dark-matter transfer function \cite{PEEBLES82}.

In Fig.(5), we show power-spectra for the density perturbation,
\begin{equation}
{\cal P}_\delta(\tau, k) \equiv { k^3 \over 2 \pi^2 } \int d^3x
\; e^{-i \vec k \cdot \vec x}  < \delta(\tau, x) \delta(\tau, 0) > \, ,
\end{equation}
arising from power-law inflation.
The bold line depicts the power-spectrum
\begin{equation}
{\cal P}^{(obs)}_\delta(k) =
{ 2 \over \pi } \, {\rm sin}\left ( { \pi \gamma \over 2 } \right ) \,
\Gamma(2- \gamma) \left(kr_0 \right)^\gamma
\end{equation}
for the observed galaxy-galaxy correlation function
$\xi_{gg}(r)= (r/ r_0)^{-\gamma}$  where $r_0= 10$ Mpc,
$\gamma=1.8$, and $\Gamma$ is the gamma function.
The corresponding biasing parameter $b_\rho$ is found by computing the
mass fluctuation on a scale of 16 Mpc:
\begin{equation}
< \left ( { \Delta M \over M}(r=16 Mpc) \right )^2 > \, = 1/ b_\rho^2 \, .
\end{equation}
For $n_s=1, 0.95, 0.9, 0.85, 0.8, 0.7, 0.5$, we compute
the biasing parameter to be $b_\rho = 0.82, 1.06, 1.34, 1.65, 2.0, 2.88,
5.6$. In order to be consistent with the biased galaxy formation,
we insist that $b_\rho < 2$  and $0.8 <  n_s < 1 $
otherwise there are not enough
fluctuations to seed galaxies. As a result, for power-law inflation
no more than 50\% of COBE's signal can arise from gravitational waves.

Previously, it had been suggested that power-law inflation with $n_s=0.5$
\cite{LLS92}  could account for the APM (Automatic Plate
machine) survey \cite{APM} which demonstrated more power than predicted
by the standard cold-dark-matter model. However, Salopek
\cite{S92} pointed out that the
production of gravitational radiation, which was neglected in
the previous calculation, could be quite significant for power-law inflation.
He was able to rule out this promising model for large scale power
since it is essential that $n_s > 0.8$. (For a careful discussion
of statistical limits on the spectral index $n_s$ using only COBE data,
consult Kurki-Suonio and Mathews \cite {Mathews94}).

\section{Conclusions}

Hamilton-Jacobi methods are a cornerstone of modern
theoretical physics, and they may be profitably applied
to numerous problems in cosmology. For example, they have
been successfully employed in deriving the Zel'dovich
approximation and its higher generalizations from
general relativity \cite {CPSS94} --\cite {SSC94}. Various researchers have
employed HJ methods in an attempt to recover the inflaton potential
from cosmological observations \cite{COPELAND93}. Moreover,
they can be used to construct inflationary models that
yield non-Gaussian primordial fluctuations \cite{SB1}; such models
could possibly resolve the problem of large scale structure
\cite{Mosc93}.  Here we have given
a careful and detailed computation of the galaxy-galaxy correlation function
and large-angle microwave background fluctuations arising
from power-law inflation, which is the most interesting
model involving gravitational radiation. We find that
the resulting spectral index for scalar perturbations must satisfy
$ n_s > 0.8$, otherwise the production of gravitational
radiation is excessive, and there are not enough fluctuations
to seed galaxies.

Our analysis is greatly facilitated
by the fact that a choice of the time-hypersurface
is not required in the computation of the probability functional
during the inflationary epoch.
Field-space diagrams are useful in visualizing a
hypersurface transformation.  However in the end when
one compares with observations one typically assumes a particular
choice of gauge. For large-angle microwave background
fluctuations which are computed using the Sachs-Wolfe
formula, comoving synchronous gauge is preferred.

Our line integral formulation of the HJ equation
(\ref{IHJE}) goes a long way in illuminating the role of time in
semi-classical general relativity. Different time-hypersurface choices
correspond to different choices of contours in superspace.
Provided spatial gauge invariance is maintained,
they all yield the same result for the generating functional.

A complete quantum formulation of the gravitational
field is still lacking. String theory is a possible
candidate, and its applications to cosmology are currently
being investigated \cite{Gasp91}, \cite{Devega93}. Our aim is more
modest in that we have restricted ourselves to the semi-classical theory
of Einstein gravity which is nonetheless
adequate in describing various quantum gravitational
phenomena including graviton fluctuations beginning initially
in the ground state \cite{S92}. We hence follow in spirit the historical
development of the theory of atomic spectra. Before
the development of the quantum theory in 1926,
the semi-classical theory of Bohr and Sommerfeld provided
a useful although imperfect description of various atoms.

\acknowledgments

D.S.S. thanks A. Barvinsky for useful discussions, and
acknowledges partial support from the Natural Sciences and
Engineering Research Council of Canada, and the Canadian Institute for
Theoretical Astrophysics in Edmonton.

\vfill\eject

\section{Figure Captions}

\noindent
{\bf Fig.(1)}: In the long-wavelength limit, the evolution
of the scalar field and the metric are shown for
inflation with an exponential potential.
For each spatial point corresponds a trajectory
(which is a straight line).
Slicing this diagram in a particular
direction would represent making a time-hypersurface choice.
$\Delta \phi(\alpha) $ denotes the scalar field fluctuation on a
time-hypersurface of uniform $\alpha= (\ln \gamma)/6$.
$\Delta \alpha(\phi)$ refers to  the metric fluctuation on a
time-hypersurface of uniform $\phi$. These perturbations are related to each
other through the slope of the $\alpha - \phi$ trajectories.

\noindent
{\bf Fig.(2)}: In the long-wavelength limit, the evolution
of the dust field and the metric are shown when
a cosmological constant is present.
$\Delta \alpha(\chi)$ refers to  the metric fluctuation on a
time-hypersurface of uniform $\chi$ (comoving, synchronous gauge).
$\Delta \chi(\alpha)$ denotes fluctuation of the dust field on a
time-hypersurface of uniform $\alpha$.
The hypersurface transformation
relating the two is more complicated than in Fig.(1)
because here the trajectories are curved.

\noindent
{\bf Fig.(3)}: The gravity wave contribution to large angle $\Delta T/ T$
can be dominant for power-law inflation which employs an exponential
potential,
$V(\phi) = V_0 \, {\rm exp} [ - \sqrt{2 \over p}  \phi ]$.
As a function of the spherical
harmonic $\ell$, the relative contribution of the
gravity waves $< \Delta T_{\ell}^2>_t$ to the total contribution
$< \Delta T_{\ell}^2>_s+ < \Delta T_{\ell}^2>_t$ is plotted for various
potential parameters $n_s= 1- 2/(p-1)$. If one normalizes to
COBE, then $n_s > 0.8$ is required to give fluctuations large enough
to produce galaxies.

\noindent
{\bf Fig.(4)}: Primordial scalar perturbations of the metric
are described by the function $\zeta$. The fluctuation
spectra for zeta are shown
for various choices of the the spectral index $n_s$
arising from power-law inflation. They have been normalized using
COBE's 2-yr data set.

\noindent
{\bf Fig.(5)}: For the present epoch, the power spectra
for the linear density perturbation $\delta$ in comoving synchronous
gauge are shown. The dark line depicts the observed two-point correlation
function describing galaxy clustering. If there is no biasing,
$n_s=0.9$ gives a good fit
to the observed data near $k=10^{-1}Mpc^{-1}$. In order that there
be enough fluctuations to seed galaxies, one
requires that the biasing parameter $b_\rho$ be less than 2
which implies that $0.8 < n_s < 1$.  As a result, for power-law
inflation, at most 50\% of COBE's signal may arise from gravitational
waves.

%
%
\begin{titlepage}
  \begin{figure}
     \begin{center}
\setlength{\unitlength}{0.240900pt}
\ifx\plotpoint\undefined\newsavebox{\plotpoint}\fi
\sbox{\plotpoint}{\rule[-0.175pt]{0.350pt}{0.350pt}}%


  \end{center}
 \end{figure}
\noindent
\centerline{Fig.(5)}
\end{titlepage}
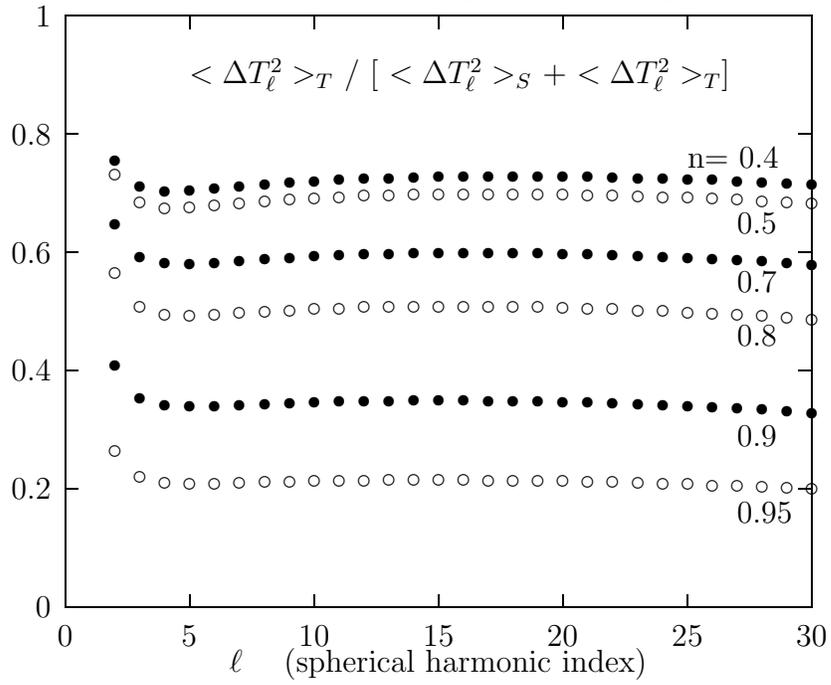
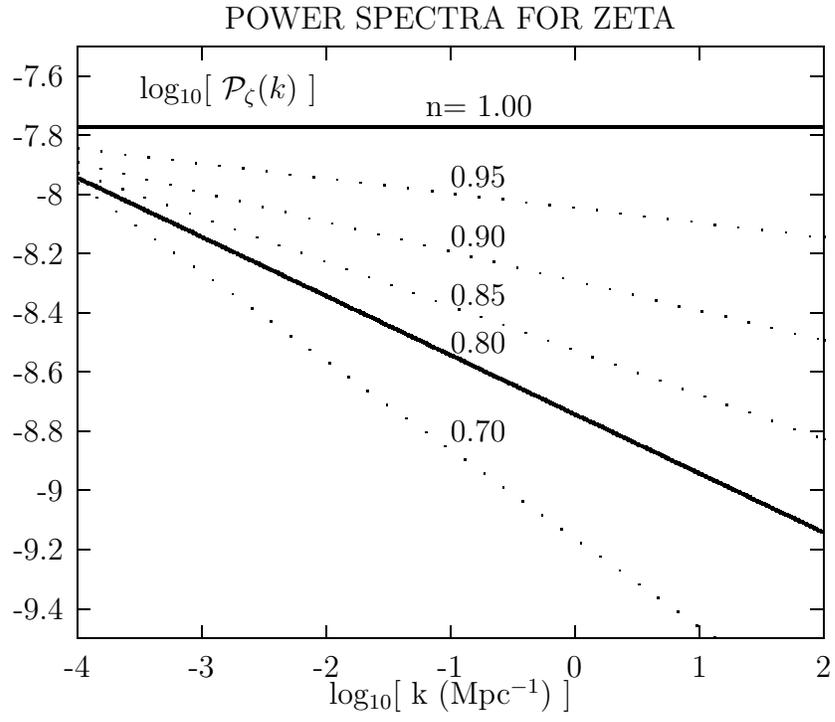
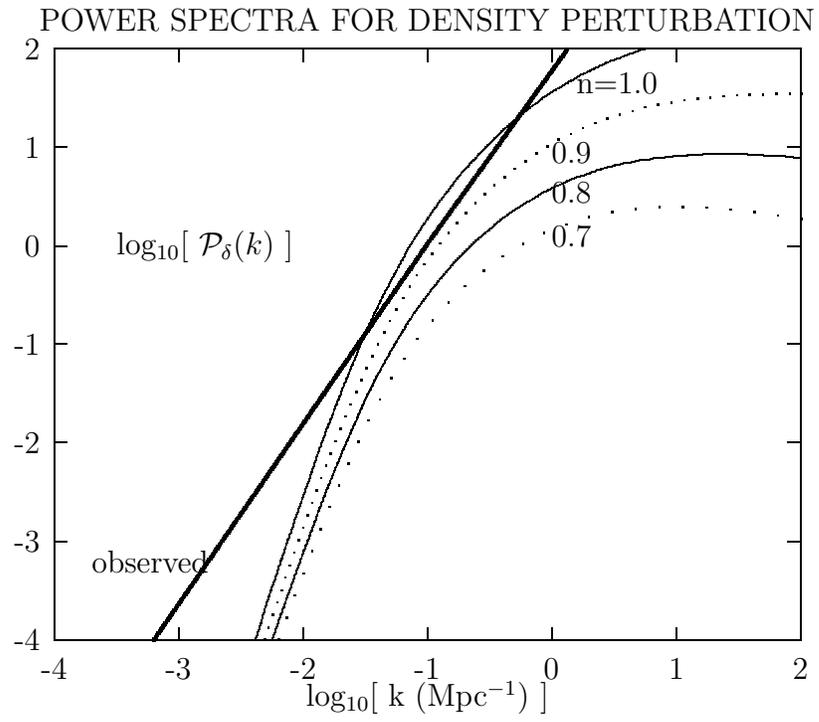
\end{document}